\documentclass[acmsmall,screen,nonacm]{acmart}
\setcopyright{none}

\AtBeginDocument{%
  }

\usepackage{microtype}
\usepackage{hyperref}
\usepackage{xcolor}
\usepackage{listings}
\usepackage[inline]{enumitem}

\usepackage{amsmath, amssymb, amsthm, amstext}
\usepackage{algorithm, algpseudocode, algorithmicx}
\usepackage{booktabs}
\usepackage{tikz}
\usetikzlibrary{positioning, arrows.meta, shapes}
\usepackage{listings}
\usepackage{xspace}
\usepackage{fvextra}
\usepackage{upquote}
\newfloat{listing}{tbp}{lol}
\floatname{listing}{Listing}
\input{code/highlight-style.tex}
\usepackage{mathpartir}
\usepackage{subcaption}
\usepackage{wrapfig}
\usepackage{wasysym}
\AtBeginDocument{
\setlength{\textfloatsep}{8pt plus 2pt minus 2pt}
\setlength{\intextsep}{8pt plus 2pt minus 2pt}
\setlength{\floatsep}{6pt plus 2pt minus 2pt}
\setlength{\dbltextfloatsep}{8pt plus 2pt minus 2pt}
\setlength{\dblfloatsep}{6pt plus 2pt minus 2pt}
}

\algrenewcommand\algorithmicprocedure{\textbf{function}}
\algrenewcommand\algorithmicend{\textbf{end}}
\algnewcommand\Match{\textbf{match}}
\algnewcommand\EndMatch{\textbf{end match}}
\algnewcommand\Case{\textbf{case}}
\algnewcommand\EndCase{\textbf{end case}}

\lstdefinelanguage{pseudo}{
  morekeywords={Input,Output,Phase,for,return,if,else,while,do,repeat,until,break,continue},
  sensitive=false,
  morecomment=[l]{//},
  morecomment=[s]{/*}{*/},
  morestring=[b]",
}

\newcommand{\tool}{\textsf{UnsafeChecker}\xspace}

\newcommand{\D}{ownership domain\xspace}
\newcommand{\DD}{lifecycle domain\xspace}
\newcommand{\DDD}{layout domain\xspace}

\begin{document}

\title{UnsafeChecker: Finding Soundness Bugs in Rust Safe Abstractions}

\author{Xizhe Yin}
\authornote{Both authors contributed equally to this research.}
\email{xizheyin@smail.nju.edu.cn}
\orcid{0000-0002-9588-8393}
\affiliation{%
  \institution{Nanjing University}
  \city{Nanjing}
  \state{Jiangsu}
  \country{China}
}

\author{Yaokun Zhang}
\authornotemark[1]
\email{yaokunzhang@smail.nju.edu.cn}
\orcid{0000-0002-7477-3642}
\affiliation{%
  \institution{Nanjing University}
  \city{Nanjing}
  \state{Jiangsu}
  \country{China}
}

\author{Yang Feng}
\orcid{0000-0002-7477-3642}
\affiliation{%
  \institution{Nanjing University}
  \city{Nanjing}
  \country{China}
}
\email{fengyang@nju.edu.cn}
\authornote{Yang Feng is the corresponding author.}

\author{Baowen Xu}
\orcid{0000-0001-7743-1296}
\affiliation{%
  \institution{Nanjing University}
  \city{Nanjing}
  \country{China}
}
\email{bwxu@nju.edu.cn}

\begin{abstract}
Rust guarantees memory safety without garbage collection through a strict ownership and borrowing system.
However, for low-level systems programming,
many widely used libraries rely on the \texttt{unsafe} keyword.
These libraries encapsulate raw-pointer operations behind safe APIs to form \emph{safe abstractions}.
A single mistake in this internal \texttt{unsafe} code can break its safety contract,
rendering the abstraction \emph{unsound} and allowing safe clients to trigger undefined behavior.
Detecting these potential soundness violations is challenging.
Existing static analysis tools for C/C++ ignore Rust-specific safety contracts,
while current Rust tools lack the deep semantic modeling required to track the contexts that raw pointers erase.

To address this gap, we present \tool,
a compiler-integrated static analysis framework for detecting potential soundness violations
in Rust safe abstractions.
\tool analyzes Rust MIR using a flow-sensitive abstract interpretation that maintains
a shared state with three components:
ownership, object validity, and layout.
Each warning rule consumes the subset of facts needed for the corresponding Rust safety obligation.
\tool reports both instruction-level undefined behavior and boundary-level contract violations
that may escape through safe APIs.
We evaluate \tool on a benchmark of 46 RustSec vulnerabilities, which contain 53 ground-truth bugs.
\tool outperforms several state-of-the-art tools, detecting 32 CVEs and covering 36 bugs (67.9\% recall) with 51.6\% alert-level precision.
Furthermore, in a large-scale scan of real-world crates on \texttt{crates.io},
\tool uncovered 114 previously unknown bugs across 83 crates, with 45 confirmed and 27 already fixed by maintainers.
\end{abstract}

\maketitle

\section{Introduction}
\label{sec:intro}

Rust has become popular in systems programming,
adopted by major projects such as Android and the Linux kernel~\cite{memory-safety-xu2021, GoogleRustAndroid2021, RustLinux2021}.
A key reason for this popularity is that Rust guarantees memory safety without garbage collection.
This makes it an ideal choice for building low-level systems that demand both high performance and strict safety.
To achieve this, Rust relies on a strict ownership and borrowing system.
For safe references, the compiler statically tracks their semantic context---specifically ownership relations, lifetimes, and type layouts
---to enforce critical \emph{safety invariants}.
This ensures that every value has a unique owner,
no reference outlives its data,
and memory accesses are strictly bounded.
As long as code remains within the safe subset of the language (\emph{Safe Rust}),
the compiler guarantees that these invariants hold,
ensuring memory safety~\cite{rustbook:book, rust-reference}.

However, the strict ownership model can be too restrictive for low-level tasks,
such as direct hardware manipulation and high-performance data structures.
To support this, Rust provides the \texttt{unsafe} keyword,
which relaxes compiler checks and permits the use of \emph{raw pointers}~\cite{rustbook:book}.
Raw pointers behave as plain addresses and do not carry the ownership, lifetime,
or layout guarantees that Safe Rust relies on.
As a result, the compiler cannot statically enforce these safety invariants for them.
In practice, library developers encapsulate \texttt{unsafe} blocks behind safe APIs,
forming \emph{safe abstractions}.
For example, the standard library implements types such as \texttt{Vec} and \texttt{Mutex} with unsafe code,
but exposes safe interfaces.
This design relies on a critical contract:
the internal \texttt{unsafe} code must manually uphold the same \emph{safety invariants} mandated by Safe Rust.
If this contract is broken, the abstraction is \emph{unsound},
allowing safe client code to trigger undefined behavior~\cite{rustrepo:misc, HowUseUnsafe2020}.
Because the compiler does not validate these conditions inside \texttt{unsafe} blocks,
a single implementation error can compromise the safety of the entire library~\cite{memory-safety-xu2021}.
This gap motivates automated analyses for detecting potential soundness violations in safe abstractions.

Existing approaches fall short because they prioritize \emph{execution errors}
over whether unsafe code upholds the contract relied on by safe APIs.
Tools designed for C and C++ operate on a flat memory model.
They check for invalid memory accesses (e.g., buffer overflows) but ignore Rust's safety contracts.
They also lack a language-level definition of the safety invariants that Safe Rust enforces.
In Rust, these invariants are part of the soundness contract of safe abstractions~\cite{rust-reference}.
For example, duplicating a raw pointer is valid in C.
In Rust, however, if that pointer represents a unique resource (e.g., \texttt{Box<T>}),
duplicating it violates ownership uniqueness.
Since C/C++ tools do not model ownership semantics, they cannot detect such contract violations.
Similarly, existing Rust tools often rely on syntactic pattern matching or general memory safety checks.
While effective for specific bug classes,
they lack the deep semantic modeling required to check that an abstraction maintains its safety invariants.
They do not track the semantic context---ownership, lifecycle, and layout---that the compiler erases in \texttt{unsafe} blocks.
Consequently, they cannot reliably check that an unsafe implementation preserves the invariants required by its safe API.

% \paragraph{Our Approach.}
To detect potential soundness violations in safe abstractions, we design \tool,
a \textbf{three-component static analysis framework} for Rust.
\tool targets bugs where unsafe pointer operations violate the safety obligations
that safe clients rely on.
It analyzes Rust MIR by tracking three kinds of facts:
ownership, object validity, and layout (Section~\ref{sec:approach}).
Specifically, the ownership component captures aliasing and resource uniqueness;
the object-validity component tracks states such as live, moved, dropped, or uninitialized;
and the layout component models pointer bounds, alignment, mutability, and offsets.
These facts are maintained in one shared flow-sensitive state.
Some checks are domain-local, such as layout-centered bounds checks,
while others combine facts across components through shared object and pointer identities.
The analysis is flow-sensitive.
It propagates this composite state along the CFG until reaching a fixpoint.
\tool then checks instruction-level UB (e.g., out-of-bounds and invalid dereference)
and boundary-level violations that can escape through safe APIs (e.g., returning a dangling pointer).

% \paragraph{Contributions.}
The contributions of this paper can be summarized as follows:
\begin{itemize}[leftmargin=*]
% \item \textbf{Integrated Analysis:}
\item \textbf{Framework:}
We propose a MIR-level static analysis that maintains ownership, object-validity,
and layout facts in a shared flow-sensitive state.
This design supports both domain-local checks, such as spatial layout violations,
and composite checks, such as dangling pointers and double frees that require
ownership and validity facts.

% \item \textbf{Compiler Integrated Tool:}
\smallskip
\item \textbf{Tool:}
We implemented \tool as a compiler plugin.
It analyzes Rust MIR efficiently and fits into existing build pipelines.

\smallskip
% \item \textbf{Evaluation:}
\item \textbf{Study:}
We evaluated \tool on a benchmark of 46 RustSec vulnerabilities
(comprising 53 ground-truth bugs).
The results show that \tool detects 32 vulnerabilities (69.6\% recall)
and covers 36 bugs (67.9\% recall),
with an alert-level precision of 51.6\%,
outperforming several state-of-the-art tools~\cite{MIRChecker2021, rudra2021, SafeDrop2023}.
Furthermore, when applied to a large-scale scan of real-world Rust crates, \tool uncovered 114 previously unknown bugs across 83 crates. Among these, developers have confirmed 45 as genuine vulnerabilities, and 27 have already been fixed.
\end{itemize}

\section{Background and Motivating Examples}
\label{sec:motivating}

This section motivates our analysis goal and the semantic gaps introduced by \texttt{unsafe}.
We first define soundness for safe abstractions and the safety invariants that Safe Rust enforces.
We then walk through three real-world bugs from our evaluation to show how unsafe implementations violate these invariants.
These examples motivate tracking three kinds of semantic context that raw pointers erase.

\subsection{Background: Soundness and Semantic Gaps}
\label{subsec:insights}

In Rust, a \emph{safe abstraction} is a public API that exposes a safe interface
while internally using \texttt{unsafe} code.
It acts as a contract: as long as the API is used from Safe Rust,
the abstraction guarantees that no undefined behavior (UB) occurs.
Because the compiler does not validate safety requirements inside \texttt{unsafe},
this contract depends on the developer upholding Rust's safety invariants in the implementation.
If the implementation violates these invariants, the abstraction becomes \emph{unsound} and safe client code may trigger UB~\cite{rust-reference, rustbook:book}.

A \emph{soundness violation} means that a safe abstraction fails to uphold this safety contract.
The Rust Reference defines soundness as follows~\cite{rust-reference}:
\begin{quote}
\emph{``Unsafe code that satisfies this property for any safe client is called sound; if unsafe code can be misused by safe code to exhibit undefined behavior, it is unsound.''}
\end{quote}
This means that there exists a \emph{safe} usage of the API that can lead to
undefined behavior (UB).
Unsoundness can manifest in two ways.
First, the unsafe implementation may directly execute an \emph{instruction-level UB}
(e.g., invalid dereference, out-of-bounds access, or use of uninitialized memory)
under safe inputs.
Second, the API may return or leak an \emph{invalid state} to safe code
(e.g., a dangling pointer or duplicated ownership),
so that UB may occur later in the caller, even if the callee itself does not crash.
We refer to the conditions that safe code relies on---
such as aliasing, validity, and layout requirements---as \emph{safety invariants}.
Ensuring soundness requires checking whether unsafe code maintains these invariants, not just detecting crashes.

Safe Rust establishes these safety invariants through three core mechanisms:
(1) \emph{unique ownership}, which ensures resources have a single owner to prevent double-free;
(2) \emph{borrowing rules}, which ensure references do not outlive the data they refer to,
preventing dangling pointers;
and (3) \emph{memory layout}, which ensures pointers are aligned and within bounds.
In this work, we categorize the corresponding safety invariants into three dimensions:
\emph{ownership}, \emph{lifecycle}, and \emph{layout}.
Within \texttt{unsafe} blocks, the compiler no longer enforces these rules,
so an analysis must check them to reason about soundness.

\subsection{Motivating Examples}
\label{subsec:examples}

We now illustrate how soundness violations manifest in real-world code,
using bugs detected in our evaluation (Section~\ref{sec:results}).

\begin{listing}[htbp]
\begin{Verbatim}[commandchars=\\\{\},numbersep=2pt, fontsize=\footnotesize, numbers=left]
\PYG{k}{pub}\PYG{+w}{ }\PYG{k}{trait}\PYG{+w}{ }\PYG{n}{ProcessMemory}\PYG{+w}{ }\PYG{p}{\PYGZob{}}
\PYG{+w}{    }\PYG{k}{fn}\PYG{+w}{ }\PYG{n+nf}{copy\PYGZus{}struct}\PYG{o}{\PYGZlt{}}\PYG{n}{T}\PYG{o}{\PYGZgt{}}\PYG{p}{(}\PYG{o}{\PYGZam{}}\PYG{n+nb+bp}{self}\PYG{p}{,}\PYG{+w}{ }\PYG{n}{addr}\PYG{p}{:}\PYG{+w}{ }\PYG{k+kt}{usize}\PYG{p}{)}\PYG{+w}{ }\PYG{p}{\PYGZhy{}\PYGZgt{}}\PYG{+w}{ }\PYG{n+nb}{Result}\PYG{o}{\PYGZlt{}}\PYG{n}{T}\PYG{p}{,}\PYG{+w}{ }\PYG{n}{Error}\PYG{o}{\PYGZgt{}}\PYG{+w}{ }\PYG{p}{\PYGZob{}}
\PYG{+w}{        }\PYG{k+kd}{let}\PYG{+w}{ }\PYG{k}{mut}\PYG{+w}{ }\PYG{n}{data}\PYG{+w}{ }\PYG{o}{=}\PYG{+w}{ }\PYG{n+nf+fm}{vec!}\PYG{p}{[}\PYG{l+m+mi}{0}\PYG{p}{;}\PYG{+w}{ }\PYG{n}{std}\PYG{p}{::}\PYG{n}{mem}\PYG{p}{::}\PYG{n}{size\PYGZus{}of}\PYG{p}{:}\PYG{p}{:}\PYG{o}{\PYGZlt{}}\PYG{n}{T}\PYG{o}{\PYGZgt{}}\PYG{p}{(}\PYG{p}{)}\PYG{p}{]}\PYG{p}{;}
\PYG{+w}{        }\PYG{n+nb+bp}{self}\PYG{p}{.}\PYG{n}{read}\PYG{p}{(}\PYG{n}{addr}\PYG{p}{,}\PYG{+w}{ }\PYG{o}{\PYGZam{}}\PYG{k}{mut}\PYG{+w}{ }\PYG{n}{data}\PYG{p}{)}\PYG{o}{?}\PYG{p}{;}
\PYG{+w}{        }\PYG{n+nb}{Ok}\PYG{p}{(}\PYG{k}{unsafe}\PYG{+w}{ }\PYG{p}{\PYGZob{}}\PYG{+w}{ }\PYG{n}{std}\PYG{p}{::}\PYG{n}{ptr}\PYG{p}{::}\PYG{n}{read}\PYG{p}{(}\PYG{n}{data}\PYG{p}{.}\PYG{n}{as\PYGZus{}ptr}\PYG{p}{(}\PYG{p}{)}\PYG{+w}{ }\PYG{k}{as}\PYG{+w}{ }\PYG{o}{*}\PYG{k}{const}\PYG{+w}{ }\PYG{n}{T}\PYG{p}{)}\PYG{+w}{ }\PYG{p}{\PYGZcb{}}\PYG{p}{)}
\PYG{+w}{    }\PYG{p}{\PYGZcb{}}
\PYG{p}{\PYGZcb{}}
\end{Verbatim}

\caption{Soundness violation via \texttt{ptr::read} in \texttt{remoteprocess}}
\label{lst:motiv1}
\end{listing}

\smallskip
\noindent
\textbf{Example A.}
Listing~\ref{lst:motiv1} shows a soundness violation in the \texttt{remoteprocess} crate.
The function \texttt{copy\_struct}, provided by the \texttt{ProcessMemory} trait,
is a \textbf{safe abstraction} that claims to safely read a struct \texttt{T} from a target process.
It encapsulates low-level memory manipulation within an \texttt{unsafe} block.
The function
(1) allocates a local byte buffer (\texttt{data});
(2) fills it with raw bytes from the target;
and (3) uses \texttt{ptr::read} to bitwise copy the buffer into a value of type \texttt{T}.

This implementation is unsound for two reasons.
\textit{First, it violates memory layout requirements, specifically pointer alignment.}
In Rust, \texttt{Vec<u8>} provides only 1-byte alignment for its buffer.
If the generic type \texttt{T} requires stricter alignment (e.g., 8 bytes for \texttt{u64}),
casting \texttt{*const u8} to \texttt{*const T} yields a pointer that is not guaranteed to be properly aligned.
Using \texttt{ptr::read} through such a pointer is undefined behavior.
\textit{Second, it violates the unique ownership guarantee.}
Rust marks types safe for bitwise duplication as \texttt{Copy} (e.g., \texttt{i32}).
Non-\texttt{Copy} types, however, often own heap allocations (e.g., \texttt{String}).
Bitwise copying such types duplicates their internal pointers,
causing two values to own the same resource.
Without a \texttt{Copy} bound on \texttt{T},
\texttt{ptr::read} creates a second owner.
This results in a double-free error
when both the returned object and the original resource
go out of scope and attempt to deallocate the same memory.

\begin{Verbatim}[commandchars=\\\{\},numbersep=2pt, fontsize=\footnotesize]
\PYG{k}{use}\PYG{+w}{ }\PYG{n}{remoteprocess}\PYG{p}{:}\PYG{p}{:}\PYG{p}{\PYGZob{}}\PYG{n}{LocalProcess}\PYG{p}{,}\PYG{+w}{ }\PYG{n}{ProcessMemory}\PYG{p}{\PYGZcb{}}\PYG{p}{;}
\PYG{c+cp}{\PYGZsh{}[}\PYG{c+cp}{derive(Debug)}\PYG{c+cp}{]}
\PYG{k}{struct}\PYG{+w}{ }\PYG{n+nc}{Victim}\PYG{p}{(}\PYG{n+nb}{String}\PYG{p}{)}\PYG{p}{;}
\PYG{k}{fn}\PYG{+w}{ }\PYG{n+nf}{main}\PYG{p}{(}\PYG{p}{)}\PYG{+w}{ }\PYG{p}{\PYGZob{}}
\PYG{+w}{    }\PYG{k+kd}{let}\PYG{+w}{ }\PYG{n}{victim}\PYG{+w}{ }\PYG{o}{=}\PYG{+w}{ }\PYG{n}{Victim}\PYG{p}{(}\PYG{l+s}{\PYGZdq{}}\PYG{l+s}{Hello from heap}\PYG{l+s}{\PYGZdq{}}\PYG{p}{.}\PYG{n}{to\PYGZus{}string}\PYG{p}{(}\PYG{p}{)}\PYG{p}{)}\PYG{p}{;}
\PYG{+w}{    }\PYG{k+kd}{let}\PYG{+w}{ }\PYG{n}{\PYGZus{}copy}\PYG{p}{:}\PYG{+w}{ }\PYG{n+nc}{Victim}\PYG{+w}{ }\PYG{o}{=}\PYG{+w}{ }\PYG{n}{LocalProcess}\PYG{p}{.}\PYG{n}{copy\PYGZus{}struct}\PYG{p}{(}\PYG{o}{\PYGZam{}}\PYG{n}{victim}\PYG{+w}{ }\PYG{k}{as}\PYG{+w}{ }\PYG{o}{*}\PYG{k}{const}\PYG{+w}{ }\PYG{n}{\PYGZus{}}\PYG{+w}{ }\PYG{k}{as}\PYG{+w}{ }\PYG{k+kt}{usize}\PYG{p}{)}\PYG{p}{.}\PYG{n}{unwrap}\PYG{p}{(}\PYG{p}{)}\PYG{p}{;}
\PYG{p}{\PYGZcb{}}
\end{Verbatim}

\noindent

A single safe client can trigger both issues.
The Proof-of-Concept (PoC) above calls \texttt{copy\_struct} on a non-\texttt{Copy} type.
It triggers UB in two ways.
First, \texttt{ptr::read::<Victim>} reads from a \texttt{Vec<u8>} buffer
that may not satisfy \texttt{Victim}'s alignment.
Second, since \texttt{Victim} is non-\texttt{Copy} and owns a \texttt{String},
the bitwise copy duplicates the owning pointer.
Dropping the value causes a double free.
This demonstrates that detecting unsoundness requires more than local crash checks;
it necessitates tracking \textbf{layout constraints} and \textbf{ownership relations}.

\begin{listing}[htbp]
\begin{Verbatim}[commandchars=\\\{\},numbersep=2pt, fontsize=\footnotesize, numbers=left]
\PYG{k}{impl}\PYG{o}{\PYGZlt{}}\PYG{n}{V}\PYG{o}{\PYGZgt{}}\PYG{+w}{ }\PYG{n+nb}{Iterator}\PYG{+w}{ }\PYG{k}{for}\PYG{+w}{ }\PYG{n}{Keys}\PYG{o}{\PYGZlt{}}\PYG{n}{V}\PYG{o}{\PYGZgt{}}\PYG{+w}{ }\PYG{p}{\PYGZob{}}
\PYG{+w}{    }\PYG{k}{fn}\PYG{+w}{ }\PYG{n+nf}{next}\PYG{p}{(}\PYG{o}{\PYGZam{}}\PYG{k}{mut}\PYG{+w}{ }\PYG{n+nb+bp}{self}\PYG{p}{)}\PYG{+w}{ }\PYG{p}{\PYGZhy{}\PYGZgt{}}\PYG{+w}{ }\PYG{n+nb}{Option}\PYG{o}{\PYGZlt{}}\PYG{k+kt}{usize}\PYG{o}{\PYGZgt{}}\PYG{+w}{ }\PYG{p}{\PYGZob{}}
\PYG{+w}{        }\PYG{k}{while}\PYG{+w}{ }\PYG{n+nb+bp}{self}\PYG{p}{.}\PYG{n}{pos}\PYG{+w}{ }\PYG{o}{\PYGZlt{}}\PYG{+w}{ }\PYG{n+nb+bp}{self}\PYG{p}{.}\PYG{n}{max}\PYG{+w}{ }\PYG{p}{\PYGZob{}}
\PYG{+w}{            }\PYG{k+kd}{let}\PYG{+w}{ }\PYG{n}{opt}\PYG{+w}{ }\PYG{o}{=}\PYG{+w}{ }\PYG{k}{unsafe}\PYG{+w}{ }\PYG{p}{\PYGZob{}}\PYG{+w}{ }\PYG{n}{ptr}\PYG{p}{::}\PYG{n}{read}\PYG{p}{(}\PYG{n+nb+bp}{self}\PYG{p}{.}\PYG{n}{head}\PYG{p}{.}\PYG{n}{add}\PYG{p}{(}\PYG{n+nb+bp}{self}\PYG{p}{.}\PYG{n}{pos}\PYG{p}{)}\PYG{p}{)}\PYG{+w}{ }\PYG{p}{\PYGZcb{}}\PYG{p}{;}
\PYG{+w}{            }\PYG{k}{if}\PYG{+w}{ }\PYG{n}{opt}\PYG{p}{.}\PYG{n}{is\PYGZus{}some}\PYG{p}{(}\PYG{p}{)}\PYG{+w}{ }\PYG{p}{\PYGZob{}}
\PYG{+w}{                }\PYG{n+nb+bp}{self}\PYG{p}{.}\PYG{n}{pos}\PYG{+w}{ }\PYG{o}{+}\PYG{o}{=}\PYG{+w}{ }\PYG{l+m+mi}{1}\PYG{p}{;}
\PYG{+w}{                }\PYG{k}{return}\PYG{+w}{ }\PYG{n+nb}{Some}\PYG{p}{(}\PYG{n+nb+bp}{self}\PYG{p}{.}\PYG{n}{pos}\PYG{+w}{ }\PYG{o}{\PYGZhy{}}\PYG{+w}{ }\PYG{l+m+mi}{1}\PYG{p}{)}\PYG{p}{;}
\PYG{+w}{            }\PYG{p}{\PYGZcb{}}
\PYG{+w}{            }\PYG{n+nb+bp}{self}\PYG{p}{.}\PYG{n}{pos}\PYG{+w}{ }\PYG{o}{+}\PYG{o}{=}\PYG{+w}{ }\PYG{l+m+mi}{1}\PYG{p}{;}
\PYG{+w}{        }\PYG{p}{\PYGZcb{}}
\PYG{+w}{        }\PYG{n+nb}{None}
\PYG{+w}{    }\PYG{p}{\PYGZcb{}}
\PYG{p}{\PYGZcb{}}
\end{Verbatim}

\caption{Dangling pointer due to \texttt{ptr::read} in iterator}
\label{lst:motiv2}
\end{listing}

\smallskip
\noindent
\textbf{Example B.}
Listing~\ref{lst:motiv2} shows a violation in the \texttt{emap} crate.
The iterator's \texttt{next} method returns a key from the map
by using \texttt{unsafe} \texttt{ptr::read} to copy bytes from the internal buffer (line~4).
As shown in the PoC below,
after \texttt{m.keys().next()} returns, the map \texttt{m} still considers the slot occupied.
Accessing it via \texttt{m.get(0)} triggers a use-after-free.

\begin{Verbatim}[commandchars=\\\{\},numbersep=2pt, fontsize=\footnotesize]
\PYG{k+kd}{let}\PYG{+w}{ }\PYG{k}{mut}\PYG{+w}{ }\PYG{n}{m}\PYG{p}{:}\PYG{+w}{ }\PYG{n+nc}{Map}\PYG{o}{\PYGZlt{}}\PYG{n+nb}{String}\PYG{o}{\PYGZgt{}}\PYG{+w}{ }\PYG{o}{=}\PYG{+w}{ }\PYG{n}{Map}\PYG{p}{::}\PYG{n}{with\PYGZus{}capacity\PYGZus{}none}\PYG{p}{(}\PYG{l+m+mi}{1}\PYG{p}{)}\PYG{p}{;}
\PYG{n}{m}\PYG{p}{.}\PYG{n}{insert}\PYG{p}{(}\PYG{l+m+mi}{0}\PYG{p}{,}\PYG{+w}{ }\PYG{n+nb}{String}\PYG{p}{::}\PYG{n}{from}\PYG{p}{(}\PYG{l+s}{\PYGZdq{}}\PYG{l+s}{hello}\PYG{l+s}{\PYGZdq{}}\PYG{p}{)}\PYG{p}{)}\PYG{p}{;}
\PYG{k+kd}{let}\PYG{+w}{ }\PYG{n}{\PYGZus{}}\PYG{+w}{ }\PYG{o}{=}\PYG{+w}{ }\PYG{n}{m}\PYG{p}{.}\PYG{n}{keys}\PYG{p}{(}\PYG{p}{)}\PYG{p}{.}\PYG{n}{next}\PYG{p}{(}\PYG{p}{)}\PYG{p}{;}\PYG{+w}{   }\PYG{c+c1}{// ptr::read copies bytes; dropping m[0] in next()}
\PYG{k+kd}{let}\PYG{+w}{ }\PYG{n}{s}\PYG{+w}{ }\PYG{o}{=}\PYG{+w}{ }\PYG{n}{m}\PYG{p}{.}\PYG{n}{get}\PYG{p}{(}\PYG{l+m+mi}{0}\PYG{p}{)}\PYG{p}{.}\PYG{n}{unwrap}\PYG{p}{(}\PYG{p}{)}\PYG{p}{;}\PYG{+w}{ }\PYG{c+c1}{// returns dangling reference \PYGZhy{}\PYGZhy{} UAF!}
\end{Verbatim}

As an implementation of the \texttt{Iterator} trait, this method acts as a safe abstraction.
The problem is that \texttt{ptr::read} creates a bitwise copy of the value (e.g., \texttt{String})
from the buffer into a temporary variable \texttt{opt}.
Since the function returns \texttt{Option<usize>} and not the value itself, \texttt{opt} is dropped at the end of the scope.
This drop frees the underlying heap memory (e.g., the string buffer).
However, the map does not update its state to mark the slot as empty, so it retains a pointer to the now-freed memory.
The compiler cannot enforce this logical state update through raw pointers inside \texttt{unsafe}.
This violation leaves the map in an inconsistent state:
subsequent access (e.g., via \texttt{get}) triggers a use-after-free (UAF), and dropping the map causes a double free.

The key observation is that \texttt{ptr::read} duplicates ownership into a temporary variable,
which is \textit{dropped at the end of the scope}.
When this temporary variable is dropped, it invalidates the resource held by the map.
By tracking the \textbf{ownership relation} and \textbf{memory liveness state} of \texttt{opt},
we can detect the dangling pointer.

\subsection{Key Insight: Tracking Lost Semantics to Check Safety Invariants}
\label{subsec:semantic-inference}

To detect these soundness violations, we must track the semantic information erased by \texttt{unsafe}
to check safety invariants.
Our key insight is that soundness relies on maintaining invariants across three interconnected dimensions:
\emph{ownership}, \emph{lifecycle}, and \emph{layout}.
Standard value-based dataflow analysis is insufficient as it treats pointers as simple values.
While it propagates memory addresses, it fails to record ownership, liveness, or layout constraints.
Consequently, it cannot explain why \texttt{ptr::read} in Listing~\ref{lst:motiv1} duplicates ownership,
or why the map in Listing~\ref{lst:motiv2} returns a reference to freed memory.

To address this, we track these semantics via a flow-sensitive abstract state.
This state records aliasing, ownership relations, lifecycle states, and layout attributes.
We propagate this state along the control flow graph (CFG),
merging states at join points to conservatively over-approximate all execution paths.
Crucially, these dimensions interact.
For example, dropping an owner invalidates all aliases to the underlying resource.

\begin{wrapfigure}{r}{0.6\linewidth}
%\vspace{-6pt}
\centering
\includegraphics[width=0.58\linewidth]{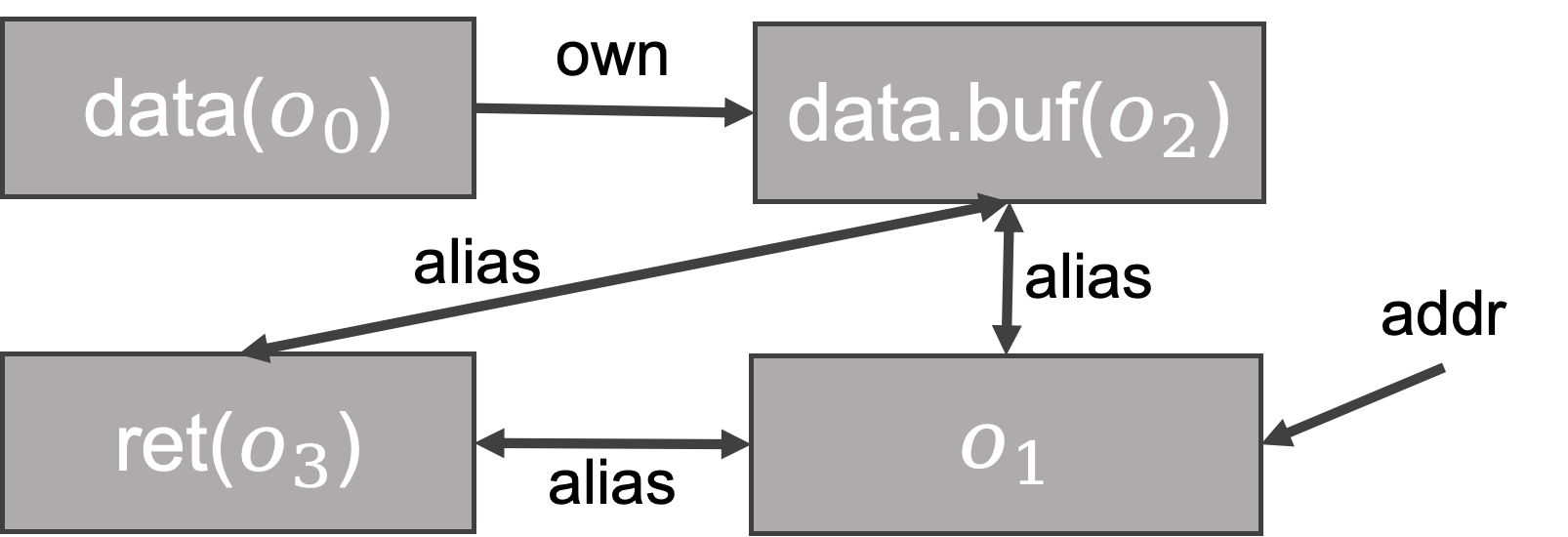}
\vspace{-8pt}
\caption{Ownership duplication in Listing~\ref{lst:motiv1}. Solid ownership
edges identify owners; alias edges connect non-owning pointer views to the
underlying resource.}
\vspace{-10pt}
\label{fig:ownership-domain}
\end{wrapfigure}

Figure~\ref{fig:ownership-domain} illustrates this using Listing~\ref{lst:motiv1}.
The figure distinguishes owner nodes from non-owning alias views:
\texttt{data(o0)} and \texttt{ret(o3)} are owning values, whereas the pointer-derived
view \texttt{o1} records an alias to the underlying buffer rather than ownership by itself.
The event order is:
first, \texttt{data(o0)} owns the heap buffer \texttt{data.buf(o2)};
second, \texttt{self.read} links the raw pointer view \texttt{o1} to that buffer;
third, \texttt{ptr::read} creates the returned owner \texttt{ret(o3)} from the same resource.
The \texttt{ptr::read} operation therefore creates a second owning view of the heap resource.
Let $\texttt{o}_0$ denote the stack object \texttt{data} (\texttt{Vec<u8>}),
and $\texttt{o}_2$ its heap buffer.
$\texttt{o}_0$ owns $\texttt{o}_2$.
The memory at \texttt{addr} is denoted as $\texttt{o}_1$.
\texttt{self.read} makes $\texttt{o}_2$ an alias of $\texttt{o}_1$.
Then, \texttt{ptr::read} creates a new object $\texttt{o}_3$ from $\texttt{o}_2$.
Our analysis tracks that $\texttt{o}_3$ aliases $\texttt{o}_2$, and by transitivity, $\texttt{o}_1$.
Thus, $\texttt{o}_1$, $\texttt{o}_2$, and $\texttt{o}_3$ form an \textit{owning alias set}.
We report an alarm when this set contains two distinct \texttt{Live} non-\texttt{Copy} owners ($\texttt{o}_1$ and $\texttt{o}_3$),
which violates the unique ownership invariant.

\section{Approach}
\label{sec:approach}

In this section, we present \tool, a flow-sensitive dataflow analysis for detecting
potential soundness violations in Rust safe abstractions.
As discussed in \S\ref{sec:motivating}, once code enters \texttt{unsafe},
raw pointers may erase semantic context that Safe Rust relies on.
\tool tracks this context with three components:
ownership, object validity, and layout.
The components are stored in one shared abstract state, but individual transfer
and warning rules update or consume only the facts relevant to the modeled MIR operation.

\subsection{Overview}
\label{subsec:overview}

\begin{figure*}[t]
    \centering
    \includegraphics[width=0.95\textwidth]{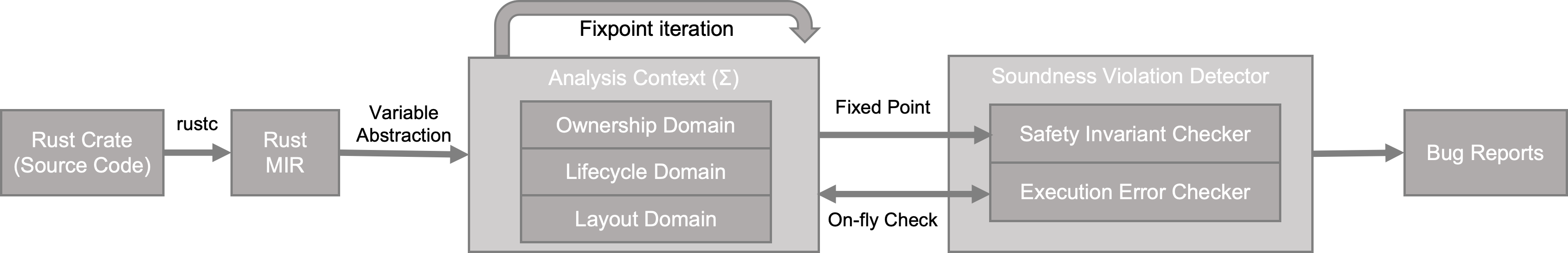}
    \caption{Overview of \tool's workflow. \tool compiles a crate to MIR, constructs a shared abstract state $\Sigma$ with ownership, object-validity, and layout facts, and iteratively applies transfer functions to a fixpoint while checking instruction-level UB and boundary-level contract violations.}
    \label{fig:workflow}
\end{figure*}

Figure~\ref{fig:workflow} illustrates the overall workflow of \tool.
Given a Rust crate, \tool first compiles it to MIR (Rust's Mid-level Intermediate Representation) using the Rust compiler, rustc.
It then abstracts MIR places into three kinds of locations:
stack owners ($\mathcal{SO}$), pointer locations ($\mathcal{P}$), and \texttt{Copy} data ($\mathcal{V}_{copy}$).
Based on this abstraction, \tool runs a flow-sensitive dataflow analysis on each function's CFG.
The analysis iteratively applies transfer functions to propagate a shared abstract state $\Sigma$
until reaching a fixpoint.
$\Sigma$ stores ownership, object-validity, and layout facts,
which capture the semantic context that raw pointers may erase.
During propagation, \tool performs instruction-level execution checks to report undefined behavior (e.g., Use-After-Free and Out-of-Bounds).
At function boundaries, it additionally checks API-level contract violations that can affect safe clients,
such as returning dangling pointers or duplicating ownership.

\paragraph{Scope of Analysis.}
Our goal is to detect potential \emph{unsoundness} in Rust safe abstractions:
whether a safe API can lead to undefined behavior when invoked from safe code.
Accordingly, we focus on undefined behavior (UB) and contract violations that arise from unsafe pointer operations
and that can compromise memory safety guarantees for safe clients.
According to the Rust Reference~\cite{rust-reference}, we target the following UB categories relevant to memory safety:
\texttt{[pointer-access]}, \texttt{[place-projection]}, \texttt{[alias]}, \texttt{[immutable]},
and the pointer- and allocation-related parts of \texttt{[invalid]}.
We also detect resource leaks, which are a critical error for many safe abstractions.
The corresponding checks consume the subset of ownership, object-validity, and layout facts
needed by each modeled Rust obligation.
In addition, we implement three lightweight auxiliary checks used in our evaluation
(Section~\ref{sec:results}).
First, we flag integer overflow in index and size computations, since wrap-around
can invalidate subsequent bounds reasoning.
Second, we flag division-by-zero, since it triggers panics and can invalidate
control-flow assumptions.
Third, we flag invalid UTF-8 passed to unsafe string constructors (e.g.,
\texttt{str::from\_utf8\_unchecked}), since it violates the documented safety
precondition of these APIs and can invalidate subsequent reasoning about string
operations.
We exclude \texttt{[race]}, \texttt{[intrinsic]}, \texttt{[target-feature]},
\texttt{[call]}, \texttt{[asm]}, \texttt{[runtime]}, and other Reference items that fall outside our scope.
This is a scoped bug-finding claim rather than a theorem of full Rust semantic soundness.
The claim is limited to the MIR operations, library/API models, and memory-safety categories
covered by \tool.
It does not cover concurrency or data races, unavailable or opaque callees,
unmodeled library behavior, inline-depth cutoffs, FFI behavior, or
pointer-integer-pointer round trips where object identity or provenance is lost.

\subsection{Variable Abstraction}
\label{subsec:variables-memory}

We model memory with \textbf{abstract locations} and \textbf{abstract heap objects}.
An abstract location $\ell \in \mathcal{L}$ represents a MIR place.
In Rust's MIR, a place is a path to a memory location, such as a local variable \texttt{x}, a field projection \texttt{x.f}, or a dereference \texttt{*p}.
To support precise field-sensitive analysis, we treat field projections (e.g., \texttt{x.f}) as distinct abstract locations from their parents (\texttt{x}).
This allows us to track partial ownership and independent borrows of disjoint fields.
However, we do not track dereferences \texttt{*p} as base locations;
instead, they are resolved to the abstract objects pointed to by \texttt{p} using the points-to relation.

We partition $\mathcal{L}$ into three disjoint subsets:
stack owners ($\mathcal{SO}$), pointer locations ($\mathcal{P}$),
and \texttt{Copy} data locations ($\mathcal{V}_{copy}$).
We also track a set of abstract heap objects $\mathcal{HO}$ that represent dynamic allocations.

\begin{itemize}
    \item $\mathcal{SO} \subseteq \mathcal{L}$:
    \textbf{stack owners}.
    Locations of non-\texttt{Copy} types (e.g., \texttt{Vec}, \texttt{Box})
    that act as roots of ownership.
    They reside on the stack and may manage heap resources.

    \item $\mathcal{P} \subseteq \mathcal{L}$:
    \textbf{pointer locations}.
    Locations holding references or raw pointers
    (e.g., \texttt{*mut T}, \texttt{\&T}).
    They access objects indirectly without owning them.
    We include raw pointers in $\mathcal{P}$ (even though they are \texttt{Copy})
    to track aliasing precisely.

    \item $\mathcal{V}_{copy} \subseteq \mathcal{L}$:
    \textbf{\texttt{Copy} data locations}.
    Other \texttt{Copy} locals (e.g., \texttt{usize}, \texttt{bool}).
    The layout domain uses these values for arithmetic reasoning.
\end{itemize}

For heap allocations, we use an \textit{allocation-site abstraction}.
All allocations from the same MIR instruction map to one canonical heap object
$h_{loc} \in \mathcal{HO}$.
Note that $\mathcal{HO} \cap \mathcal{L} = \emptyset$.
We unify all resources whose lifecycle must be tracked by defining
\textbf{ownership nodes} as $\mathcal{O} = \mathcal{SO} \cup \mathcal{HO}$.
This set includes both stack owners and heap objects.

For a composite ownership root $v \in \mathcal{SO}$ that encapsulates an internal pointer used for memory access
(e.g., a \texttt{Vec}'s buffer pointer),
we denote the corresponding pointer location (an access path in $\mathcal{L}$) by $p_v \in \mathcal{P}$.

To illustrate, consider the \texttt{Vec<T>} variable \texttt{data} in Listing~\ref{lst:motiv1}.
The variable \texttt{data} resides on the stack ($\texttt{data} \in \mathcal{SO}$)
and manages a dynamically allocated buffer ($\texttt{data.buf} \in \mathcal{HO}$).
In Safe Rust, the ownership model ensures that when \texttt{data} goes out of scope,
it automatically deallocates $\texttt{data.buf}$.
Unsafe Rust, however, allows pointers ($\mathcal{P}$) to bypass these checks,
potentially introducing aliasing, invalid lifecycle states, or out-of-bounds access.
Based on the variable abstraction above, \tool tracks these risks through three components, as detailed next.

\subsection{Ownership Domain (\texorpdfstring{$\sigma_{A}$}{sigma A})}
\label{subsec:ownership-domain}

To detect memory safety violations in \texttt{unsafe} code, \tool tracks the ownership topology that is typically erased during compilation.
The ownership domain $\mathcal{D}_{Own}$ approximates the relational structure of the heap by tracking three key relations: $\langle \mathit{Pt}, \mathit{Own}, \mathit{Alias} \rangle$.

\begin{itemize}[leftmargin=*]
    \item \textbf{Points-to Relation ($\mathit{Pt}: \mathcal{P} \rightharpoonup \wp(\mathcal{O})$).}
    A standard may-points-to map that links each pointer location $p$ to the set of objects it may reference.
    This relation resolves indirect accesses: dereferencing $p$ is treated as an access to all $o \in \mathit{Pt}(p)$.
    At control-flow joins, points-to sets are merged by union, so $\mathit{Pt}(p)$ accumulates possible targets from all incoming paths.

    \item \textbf{Ownership Hierarchy ($\mathit{Own}: \mathcal{O} \to \wp(\mathcal{HO})$).}
    Tracks the \textbf{vertical} management of resources. For an object $o$, $\mathit{Own}(o)$ is the set of heap resources for which $o$ is currently responsible.
    This explicitly represents the RAII structure (e.g., a \texttt{Vec} owning its buffer), allowing the analysis to determine which resources are affected when an object is dropped.

    \item \textbf{Ownership Aliasing ($\mathit{Alias}: \mathcal{O} \to \wp(\mathcal{O})$).}
    Tracks the \textbf{horizontal} aliasing between owners.
    $\mathit{Alias}(o)$ is the set of objects that claim ownership of the \emph{same} underlying resource as $o$.
    While Safe Rust enforces unique ownership ($|\mathit{Alias}(o)| = 1$), unsafe code can introduce duplication (e.g., via \texttt{ptr::read}).
    We track these sets to detect when multiple objects believe they own the same resource, which is the structural precursor to double-free bugs.
\end{itemize}

The domain forms a lattice ordered by component-wise subset inclusion ($\sqsubseteq$).
The join operation ($\sqcup$) computes the component-wise set union, yielding a conservative may-approximation for the modeled ownership facts.
For the aliasing component, the union accumulates ownership claims from all incoming control-flow paths:
$\mathit{Alias}_{\sqcup}(o) = \mathit{Alias}_1(o) \cup \mathit{Alias}_2(o)$.
This ensures that if an ownership duplication exists on \emph{any} path, it is preserved and checked in the subsequent analysis.

\subsection{Object-Validity (Lifecycle) Domain}
\label{subsec:lifecycle-domain}

The second dimension of our analysis is the \textbf{object-validity domain}
$\mathcal{D}_{Life}$, which tracks the runtime validity of memory objects.
While Safe Rust relies on static lifetimes to prevent use-after-free,
unsafe code bypasses these checks, necessitating a flow-sensitive model of object states.
We retain the term \emph{lifecycle} only to denote these runtime object states (e.g., \texttt{Live}, \texttt{Moved}, and \texttt{Dropped}),
distinct from Rust's compile-time \emph{lifetime} system.
We define the lifecycle state $\sigma_O: \mathcal{O} \to \wp(\Sigma_{atom})$ as a mapping from each object to a set of possible atomic states.
Elements of $\mathcal{D}_{Life}$ are such maps.

We model the lifecycle of a Rust object using a finite set of mutually exclusive atomic states $\Sigma_{atom}$.
The base state is \textbf{\texttt{Live}}, representing a fully initialized and valid object.
Memory that is allocated but not yet initialized is tracked as \textbf{\texttt{Uninit}}; reading from it is undefined behavior.
When ownership is transferred, the object enters the \textbf{\texttt{Moved}} state, and subsequent access is a Use-After-Move error.
Once a destructor runs, the object becomes \textbf{\texttt{Dropped}},
and the pointer to it is no longer valid, leading to a dangling pointer.
Any access to this pointer constitutes a Use-After-Free error.
Special handling is required for manual management:
\textbf{\texttt{ManuallyDropped}} indicates an object (e.g., wrapped in \texttt{ManuallyDrop}) that remains valid but whose automatic destructor is suppressed.
Finally, \textbf{\texttt{Forgotten}} represents an object explicitly consumed (e.g., via \texttt{mem::forget}); it is semantically moved, and its resources are leaked rather than freed.

To handle control-flow uncertainty (e.g., a variable is dropped in one branch but not another),
we construct the domain as a power set lattice $\mathcal{S} = \wp(\Sigma_{atom})$.
In this lattice, an abstract state $S \in \mathcal{S}$ represents the disjunction of all possible runtime states,
ordered by set inclusion ($\subseteq$).
At control-flow merge points, the join operation ($\sqcup$) is defined as set union ($\cup$),
which yields a conservative may-approximation for the modeled object-validity facts.
For example, if a variable is initialized on one path and moved on another,
its state at the merge point is exactly $\{\texttt{Live}, \texttt{Moved}\}$.
This power-set design is necessary for precise error reporting.
Rather than collapsing to a generic ``Unknown'' state,
\tool retains the specific set of possible states.
This supports conditional reports
(e.g., ``Potential Use-After-Free on some paths'').

\subsection{Layout Domain}
\label{subsec:layout-domain}

The third component of our analysis is the \textbf{layout domain}
$\mathcal{D}_{Lay}$,
which models the physical properties of pointers relative to their underlying memory allocations.
The \DDD complements the qualitative reasoning of \D and \DD by maintaining
byte-level precision for pointer bounds, alignment, and mutability.
This enables checks for array indexing and pointer arithmetic.

For a pointer $p$, the layout domain tracks its abstract state $\sigma_{L}(p)$ as a tuple:
\[ \sigma_{L}(p) = (\mathit{base}, \mathit{size}, \mathit{offset}, \mathit{align}, \mathit{null}, \mathit{mut}) \]
Formally, $\sigma_L$ is a map from pointer locations to such tuples.
Here, $\mathit{base}$ denotes the target memory block (heap, stack, or $\top$),
while $\mathit{size}$ represents the minimum guaranteed buffer capacity.
The $\mathit{offset}$ interval $[l, u]$ tracks the pointer's position relative to the base,
and $\mathit{align}$ records the alignment guarantee.
Finally, $\mathit{null}$ and $\mathit{mut}$ capture the pointer's nullability and mutability.

We use an explicit invalid-pointer element as the bottom state of the layout domain:
\[
    \bot_{\text{Lay}} = (\bot, 0, \bot, 1, \texttt{MaybeNull}, \texttt{Unknown})
\]

We define the partial order $\sqsubseteq_{\text{Lay}}$ so that larger elements
represent weaker safety guarantees.
Specifically, a state $s_2$ is considered less precise than $s_1$ ($s_1 \sqsubseteq_{\text{Lay}} s_2$) if it provides
a \emph{smaller} guaranteed buffer size ($z_1 \ge z_2$),
a \emph{smaller} alignment ($a_1 \ge a_2$), or a \emph{wider} offset interval ($[l_1, u_1] \subseteq [l_2, u_2]$).
For the discrete attributes, we enforce $\texttt{NonNull} \sqsubseteq \texttt{MaybeNull}$ and $\texttt{Mut} \sqsubseteq \texttt{Immut}$.
Intuitively, $\sigma_L$ stores conservative spatial facts.
It uses lower bounds for \emph{guarantees} (e.g., $\mathit{size}$ and $\mathit{align}$),
and an over-approximating interval for possible offsets.
Therefore, the join operation ($\sqcup_{\text{Lay}}$) keeps only guarantees that
hold on all incoming paths,
while merging the set of possible offsets.
It also downgrades capabilities (e.g., to \texttt{Immut} or \texttt{MaybeNull})
unless they are guaranteed everywhere.

At control-flow merge points, the join operation $\sqcup_{\text{Lay}}$
computes the safe lower bound of guarantees.
For example, it takes the minimum capacity ($\mathit{size}_{\sqcup} = \min(z_1, z_2)$)
and the union of offset intervals.
If the target bases differ ($b_1 \neq b_2$), we set $b_\sqcup = \top$
but keep the common lower-bound constraints.
This decouples spatial safety from temporal safety:
even if the exact object identity is uncertain,
\tool can still prove that an access stays within the shared physical bounds.

Finally, to support low-level pointer arithmetic and casting (e.g., reinterpreting \texttt{*mut u64} as \texttt{*mut u8}),
we track all offsets and sizes in bytes.
The $\mathit{size}$ field reflects the total allocated capacity rather than the initialized length.
Although accessing uninitialized bytes within the allocated bounds is \emph{spatially safe} (i.e., no buffer overflow),
it constitutes Undefined Behavior.
This violation is detected independently by the lifecycle domain, which tracks the \texttt{Uninit} state.

\subsection{Abstract Semantics and Dataflow Propagation}
\label{subsec:dataflow}

We formalize the static analysis as a flow-sensitive abstract interpretation over the Rust MIR.
The analysis propagates the composite abstract state $\Sigma = (\sigma_A, \sigma_O, \sigma_{L})$ through the control flow graph (CFG).
This section defines the lattice structure, the symbolic initial state for procedure summaries, and the transfer functions governing state transitions.

The global abstract domain is the product of the three sub-domains defined above:
$\mathcal{D} = \mathcal{D}_{Own} \times \mathcal{D}_{Life} \times \mathcal{D}_{Lay}$.
The composite partial order $\sqsubseteq$ and join operator $\sqcup$ are defined component-wise,
applying the respective domain-specific operations described in \S\ref{subsec:ownership-domain}--\ref{subsec:layout-domain}.
This lets the analysis compute a conservative may-approximation of the modeled program facts across all components simultaneously.

\subsubsection{Symbolic Initial State}
\label{subsubsec:initial-state}

To support the analysis of libraries, we construct a \textbf{generic symbolic precondition} $\Sigma_{\text{init}}$ at the entry points (public APIs).
For internal functions, the initial state is directly propagated from their call sites via full inlining.
For an entry point function $F$, we construct a symbolic initial state $\Sigma_{\text{init}}$ that reflects Rust's ABI guarantees:

\paragraph{Ownership.}
We assume mutable references do not alias, and owned objects form disjoint trees.
Each ownership node $o$ starts in a singleton alias set
($\mathit{Alias}(o) \leftarrow \{o\}$).
Each pointer argument $p \in Args_F$ points to a fresh symbolic object
($\mathit{Pt}(p) = \{o_p\}$).
Local pointers start empty ($\mathit{Pt}(p) = \emptyset$).
For composite arguments $v$ (e.g., \texttt{Vec}),
we create a fresh heap object $h_v$ and set $\mathit{Own}(v) = \{h_v\}$.

\paragraph{Lifecycle.}
We assume caller-provided values are valid.
Input arguments start as $\texttt{Live}$.
Local variables start as $\texttt{Uninit}$ to capture use-before-def patterns in unsafe code.

\paragraph{Layout.}
Pointers derived from arguments start with symbolic constraints.
For a container argument $v$, its internal pointer $p_v$ has a symbolic capacity $\kappa_v$
and the alignment of $T$:
\[
    \sigma_{L}(p_v) =
    (h_v, \kappa_v \cdot \text{size\_of}(T), [0, 0], \text{align\_of}(T), \texttt{NonNull}, \text{mut\_of}(v))
\]
We also set $\mathit{Pt}(p_v) = \{h_v\}$.
This representation lets \tool check constraints such as $i < \kappa_v$
without concrete values.

\subsubsection{Auxiliary Operations}
\label{subsubsec:aux-funcs}

To simplify the operational semantics,
we define auxiliary functions that update the ownership topology
and lifecycle states.
They capture common patterns such as resource destruction and overwriting.

\paragraph{Topology Updates (Transfer, Alias Merge, and Isolate)}
These operations maintain the integrity of the ownership graph during variable assignments.

\begin{itemize}[leftmargin=*]
    \item \textbf{Transfer ($\textsc{Transfer}(y, x, \sigma_A)$):} 
    Models the standard ownership transfer (move semantics).
    When a value is moved from $x$ to $y$,
    this operation updates the topology by structurally replacing $x$ with $y$ in all aliasing sets and
    transferring the heap ownership mapping from $x$ to $y$.
    This ensures that the new owner $y$ inherits the exact relationships of the original owner $x$.
    
    \item \textbf{Alias Merge ($\textsc{AliasMerge}(y, x, \sigma_A)$):} 
    Handles the creation of ownership aliases,
    such as when using \texttt{ptr::read} to duplicate a non-\texttt{Copy} value.
    This operation merges the equivalence classes of $x$ and $y$ into a single clique
    and merges heap ownership information by $\mathit{Own}(y) \leftarrow \mathit{Own}(y) \cup \mathit{Own}(x)$.
    This reflects the physical reality that both variables now claim ownership of the \emph{same} underlying resource,
    allowing the analysis to track potential double-free violations if both are dropped.

    \item \textbf{Isolate ($\textsc{Isolate}(x, \sigma_A)$):}
    Disconnects a variable $x$ from the ownership topology.
    Before $x$ is overwritten or invalidated,
    this operation removes it from all alias sets of its neighbors and resets its own alias set to a singleton.
    This preserves the symmetry of the aliasing relation and ensures that the old identity of $x$ does not affect the new value.
    It also clears $x$'s heap ownership mapping ($\mathit{Own}(x) \leftarrow \emptyset$).
\end{itemize}

\paragraph{Resource Management (Drop and Clear)}
These operations handle the destruction and overwriting of resources to prevent leaks and invalid access.

\begin{itemize}[leftmargin=*]
    \item \textbf{Recursive Drop ($\textsc{RecursiveDrop}(u, \Sigma)$):} 
    Models the cascading execution of destructors.
    It propagates the \texttt{Dropped} state not just to $u$, but to all its ownership aliases.
    \[
        \textsc{RecursiveDrop}(u, \Sigma) = \text{fold}(\Sigma', \sigma_A.\mathit{Own}(u), \lambda(\sigma, h). \textsc{RecursiveDrop}(h, \sigma))
    \]
    where $\Sigma'$ marks all aliases of $u$ as \texttt{Dropped}.
    This propagation is essential for detecting temporal violations: if one owner drops the resource,
    all other aliases must immediately reflect this invalid state to detect subsequent use-after-free errors.
    
    \item \textbf{Clear ($\textsc{Clear}(x, \Sigma)$):} 
    Prepares a variable $x$ for overwriting after MIR drop effects have been handled
    explicitly by \textsf{T-Drop}.
    If $x \in \mathcal{O}$, it freshens the abstract identity of $x$ by calling
    \textsc{Isolate} to disconnect it from the previous ownership topology.
    If $x \in \mathcal{P}$, it resets pointer-specific state by setting $\mathit{Pt}(x) \leftarrow \emptyset$ and $\sigma_L(x) \leftarrow \bot_{\text{Lay}}$.
    If $x \in \mathcal{V}_{copy}$, it performs no state mutation.
    It does not introduce an additional destructor beyond the drops already present in MIR.
\end{itemize}

\paragraph{State Transitions (Consume and Forget)}
We employ lightweight helpers to model affine type semantics:

\begin{itemize}[leftmargin=*]
    \item \textbf{Consume ($\textsc{Consume}(x, \sigma_O)$):}
    Transitions a variable $x$ and all its non-\texttt{Copy} sub-locations (fields) to the \texttt{Moved} state.
    This invalidates $x$ and any derived field pointers, preventing Use-After-Move errors.
    
    \item \textbf{Forget ($\textsc{Forget}(x, \Sigma)$):}
    Transitions $x$ to the \texttt{Forgotten} state (modeling \texttt{mem::forget}).
    Crucially, this operation propagates the \texttt{Forgotten} state to all fields of $x$,
    ensuring that pointers to sub-fields (e.g., \texttt{\&x.f}) are invalidated even if taken before the call.
\end{itemize}

\subsubsection{Transfer Functions (Operational Semantics)}
\label{subsubsec:transfer-funcs}

We formalize the transfer functions as inference rules of the form $\frac{\text{Instruction}, \quad \text{Preconditions}}{\Sigma \to \Sigma'}$. These rules are summarized in Figure~\ref{fig:transfer-rules}.
For presentation, we write a small instruction language that abstracts common MIR patterns,
including allocation (\texttt{x = Alloc(T)}),
address-taking and casts (\texttt{p = AddrOf(k, x)} and \texttt{p = Cast(k, q)}),
and pointer arithmetic (\texttt{p = q.offset(i)}).
Type parameters, \texttt{size\_of}, \texttt{align\_of}, field offsets, mutability, and ABI layout facts are obtained from rustc MIR type and layout queries.
We write $\Sigma = (\sigma_A, \sigma_O, \sigma_L)$.
When a rule invokes an auxiliary operation and obtains an intermediate state
$\Sigma_1$ (e.g., \textsc{Clear} or \textsc{Isolate}),
we write $\Sigma_1 = (\sigma_{A1}, \sigma_{O1}, \sigma_{L1})$
and apply subsequent updates to these components.
We write $\textsc{StrongTarget}(O_{dst})$ when $O_{dst}=\{o\}$ and $o$ denotes a singleton abstract object in the current context.
Allocation-site summaries and merged objects are not strong targets and therefore use weak updates.

\begin{figure*}[h]
\footnotesize
\begin{mathpar}
\inferrule*[Lab=\textbf{\textsf{T-Alloc}}]{
    \texttt{x = Alloc(T)} \\ h = \text{fresh\_heap}(x) \\ p_x \in \mathcal{P}
}{
    {
    \begin{array}{l}
        \Sigma_1 = \textsc{Clear}(x, \Sigma) \\
        \sigma_O' = \sigma_{O1}[x \mapsto \{\texttt{Live}\}, h \mapsto \{\texttt{Live}\}] \\
        \sigma_A' = \sigma_{A1}[\mathit{Own}(x) \mapsto \{h\}] \\
        \sigma_A'.\mathit{Pt}(p_x) = \{h\} \\
        \sigma_L' = \sigma_{L1}[p_x \mapsto (h, \text{size}(T), [0, 0], \text{align}(T), \texttt{NonNull}, \texttt{Mut})]
    \end{array}
    }
}

\and
\inferrule*[Lab=\textbf{\textsf{T-Move}}]{
    \texttt{y = x} \\ !Copy(x)
}{
    {
    \begin{array}{l}
        \Sigma_1 = \textsc{Clear}(y, \Sigma) \\
        \sigma_A' = \textsc{Transfer}(y, x, \sigma_{A1}) \\
        \sigma_O' = \textsc{Consume}(x, \sigma_{O1})[y \mapsto \{\texttt{Live}\}] \\
    \end{array}
    }
}
\and
\inferrule*[Lab=\textbf{\textsf{T-ReadBits}}]{
    \texttt{y = ptr::read(p)} \\ y \in \mathcal{O}
}{
    {
    \begin{array}{l}
        \Sigma_1 = \textsc{Clear}(y, \Sigma) \\
        O_{src} = \sigma_{A1}.\mathit{Pt}(p) \\
        \sigma_O' = \sigma_{O1}[y \mapsto \{\texttt{Live}\}] \\
        \sigma_A' = \text{fold}(\sigma_{A1}, O_{src}, \lambda(\sigma, o). \textsc{AliasMerge}(y, o, \sigma))
    \end{array}
    }
}
\and
\inferrule*[Lab=\textbf{\textsf{T-ManDrop}}]{
    \texttt{y = ManuallyDrop(x)}
}{
    {
    \begin{array}{l}
        \Sigma_1 = \textsc{Clear}(y, \Sigma) \quad
        \sigma_A' = \textsc{Transfer}(y, x, \sigma_{A1}) \\
        \sigma_O' = \textsc{Consume}(x, \sigma_{O1})[y \mapsto \{\texttt{ManuallyDropped}\}]
    \end{array}
    }
}
\and
\inferrule*[Lab=\textbf{\textsf{T-Load}}]{
    \texttt{y = *p} \quad y \in \mathcal{V}_{copy}
}{
    \Sigma' = \textsc{Clear}(y, \Sigma)
}
\and
\inferrule*[Lab=\textbf{\textsf{T-Store-S}}]{
    \texttt{*p = x} \\ O_{dst} = \sigma_A.\mathit{Pt}(p) \\ x \in \mathcal{O} \\ O_{dst} = \{o\} \\ \textsc{StrongTarget}(O_{dst})
}{
    {
    \begin{array}{l}
        \Sigma_1 = \textsc{Clear}(o, \Sigma) \quad
        \sigma_A' = \textsc{Transfer}(o, x, \sigma_{A1}) \\
        \sigma_O' = \textsc{Consume}(x, \sigma_{O1})[o \mapsto \{\texttt{Live}\}]
    \end{array}
    }
}
\and
\inferrule*[Lab=\textbf{\textsf{T-Store-W}}]{
    \texttt{*p = x} \\ O_{dst} = \sigma_A.\mathit{Pt}(p) \\ x \in \mathcal{O} \\ \neg \textsc{StrongTarget}(O_{dst})
}{
    {
    \begin{array}{l}
        \sigma_A' = \text{fold}(\sigma_A, O_{dst}, \lambda(\sigma, o). \textsc{AliasMerge}(o, x, \sigma)) \\
        \sigma_O' = \textsc{Consume}(x, \sigma_O)[\forall o \in O_{dst}: o \mapsto \sigma_O(o) \cup \{\texttt{Live}\}]
    \end{array}
    }
}
\and
\inferrule*[Lab=\textbf{\textsf{T-Drop}}]{
    \texttt{drop(x)}
}{
    \Sigma' = \textsc{RecursiveDrop}(x,\Sigma)
}
\and
\inferrule*[Lab=\textbf{\textsf{T-Forget}}]{
    \texttt{mem::forget(x)}
}{
    \Sigma' = \textsc{Forget}(x, \Sigma)
}
\and
\inferrule*[Lab=\textbf{\textsf{T-PtrWrite}}]{
    \texttt{ptr::write(p, x)} \\ O_{dst} = \sigma_A.\mathit{Pt}(p) \\ O_{dst} = \{o\}
}{
    {
    \begin{array}{l}
        \Sigma_1 = \textsc{Isolate}(o, \Sigma) \quad
        \sigma_A' = \textsc{Transfer}(o, x, \sigma_{A1}) \\
        \sigma_O' = \textsc{Consume}(x, \sigma_{O1})[o \mapsto \{\texttt{Live}\}]
    \end{array}
    }
}
\and
\inferrule*[Lab=\textbf{\textsf{T-AddrOf}}]{
    \texttt{p = AddrOf(k, x)}
}{
    {
    \begin{array}{l}
        \Sigma_1 = \textsc{Clear}(p, \Sigma) \quad
        \sigma_A' = \sigma_{A1}[\mathit{Pt}(p) \mapsto \{x\}] \\
        \sigma_L' = \sigma_{L1}[p \mapsto
        (x, \text{size}(x), [0, 0], \text{align}(x), \texttt{NonNull}, \text{mut}(k))]
    \end{array}
    }
}
\and
\inferrule*[Lab=\textbf{\textsf{T-Cast}}]{
    \texttt{p = Cast(k, q)} \\
    \sigma_L(q) = (b_q, s_q, o_q, a_q, n_q, m_q)
}{
    {
    \begin{array}{l}
        \Sigma_1 = \textsc{Clear}(p, \Sigma) \quad
        \sigma_A' = \sigma_{A1}[\mathit{Pt}(p) \mapsto \sigma_{A1}.\mathit{Pt}(q)] \\
        \sigma_L' = \sigma_{L1}[p \mapsto (b_q, s_q, o_q, a_q, n_q, \text{mut}(k))]
    \end{array}
    }
}
\and
\inferrule*[Lab=\textbf{\textsf{T-UsePtr}}]{
    \texttt{p = Use(q)}
}{
    {
    \begin{array}{l}
        \Sigma_1 = \textsc{Clear}(p, \Sigma) \quad
        \sigma_A' = \sigma_{A1}[\mathit{Pt}(p) \mapsto \sigma_{A1}.\mathit{Pt}(q)] \\
        \sigma_L' = \sigma_{L1}[p \mapsto \sigma_L(q)]
    \end{array}
    }
}
\and
\inferrule*[Lab=\textbf{\textsf{T-Offset}}]{
    \texttt{p = q.offset(i)} \\ \sigma_{L}(q) = (b, s, [l, u], a, n, m) \\
    \text{val}(i) = [i_{min}, i_{max}] \\ \text{sz} = \text{sizeof}(T)
}{
    {
    \begin{array}{l}
        \Sigma_1 = \textsc{Clear}(p, \Sigma) \\
        \sigma_A' = \sigma_{A1}[\mathit{Pt}(p) \mapsto \sigma_{A1}.\mathit{Pt}(q)] \\
        \sigma_L' = \sigma_{L1}[p \mapsto (b, s, [l + i_{min} \cdot \text{sz}, u + i_{max} \cdot \text{sz}], \gcd(a, \text{sz}), n, m)]
    \end{array}
    }
}
\end{mathpar}
\caption{Core Intra-Procedural Transfer Functions.}
\label{fig:transfer-rules}
\end{figure*}

\paragraph{Rule intuition.}
\textsc{T-Alloc} creates a fresh heap object and initializes all three domains:
it sets lifecycle to \texttt{Live},
records ownership edges, and initializes points-to and layout facts.
The fresh pointer starts with \texttt{Mut} capability because a newly allocated
owned object is uniquely mutable at allocation time; later borrows and casts can
downgrade or propagate this capability.
\textsc{T-Move} transfers the ownership topology and marks the source as \texttt{Moved}.
\textsc{T-ReadBits} models \texttt{ptr::read}:
it creates a new owner and merges alias sets to represent ownership duplication.
\textsc{T-Drop}, \textsc{T-Forget}, and \textsc{T-ManDrop} update lifecycle
and trigger the corresponding topology updates.
\textsc{T-ManDrop} moves the binding $x$ into the wrapper $y$:
the source binding $x$ becomes moved, but $y$ denotes valid storage with suppressed
automatic destruction.
Therefore, \texttt{ManuallyDropped} is not included in \texttt{InvalidSet};
access through the wrapper is not reported as use-after-move.
\textsc{T-Load} reads \texttt{Copy} data and does not change ownership or lifecycle.
Non-\texttt{Copy} value movement is handled by \textsc{T-Move} or by
ownership-producing API rules such as \textsc{T-ReadBits}, rather than by
\textsc{T-Load}.
\textsc{T-Store-S} and \textsc{T-Store-W} write through pointers,
using strong updates only when the destination denotes a singleton target and weak
updates for merged targets or allocation-site summaries.
\textsc{T-PtrWrite} models \texttt{ptr::write},
which overwrites memory without dropping the old value.
\textsc{T-AddrOf}, \textsc{T-Cast}, and \textsc{T-UsePtr} handle
address-taking, casts, and pointer propagation separately.
address-taking creates target and layout facts from a place, casts preserve
points-to and layout facts except for mutability changes, and pointer
propagation copies the existing facts unchanged.
\textsc{T-Offset} propagates bounds and alignment through pointer arithmetic;
the $\gcd$ term gives a conservative lower bound on the resulting pointer alignment
after advancing by multiples of $\text{sizeof}(T)$.

\subsubsection{Termination and Fixed-Point Iteration}
\label{subsubsec:termination}

The solver uses a standard worklist algorithm over the CFG.
It iterates until all program points reach a post-fixpoint.
Termination follows from the height of the abstract domains:
\begin{itemize}
    \item \textbf{Finite domains ($\sigma_A, \sigma_O$).}
    The ownership and lifecycle domains range over finite sets.
    $\mathcal{L}$ is finite for a function body,
    and $\mathcal{HO}$ is finite due to the allocation-site abstraction.
    The transfer functions are monotone w.r.t.\ $\sqsubseteq$,
    so ascending chains stabilize.

    \item \textbf{Layout domain ($\sigma_L$).}
    The interval component has infinite height.
    We apply widening at loop headers when the interval grows,
    and optionally apply narrowing based on simple loop guards
    (e.g., \texttt{i < len}) to recover precision.
\end{itemize}
This combination guarantees termination and keeps the layout analysis practical.

\begin{figure*}[h]
\footnotesize
\begin{mathpar}
\inferrule*[Lab=\textbf{\textsf{E-AliasRet}}]{
    o_1, o_2 \in \mathcal{V}_{out} \\ o_1 \neq o_2 \\ 
    \mathit{Own}(o_1) \cap \mathit{Own}(o_2) \neq \emptyset
}{
    \Sigma_{final} \vdash \iota_{\text{end}} \implies \textsc{Err(DoubleOwnership)}
}
\and
\inferrule*[Lab=\textbf{\textsf{E-MutAlias}}]{
    p_1, p_2 \in \mathcal{V}_{out} \land p_1 \neq p_2 \\
    \texttt{is\_ref}(p_1) \land \texttt{is\_ref}(p_2) \\
    \mathit{Pt}(p_1) \cap \mathit{Pt}(p_2) \neq \emptyset \\
    (\sigma_L(p_1).\mathit{mut} = \texttt{Mut} \lor \sigma_L(p_2).\mathit{mut} = \texttt{Mut})
}{
    \Sigma_{final} \vdash \iota_{\text{end}} \implies \textsc{Err(MutAlias)}
}
\and
\inferrule*[Lab=\textbf{\textsf{E-Dangling}}]{
    p \in \mathcal{V}_{out} \\ \text{sizeof}(T) > 0 \\ \exists o \in \mathit{Pt}(p). \sigma_O(o) \cap \{\texttt{Dropped}, \texttt{Moved}, \texttt{Forgotten}, \texttt{Uninit}\} \neq \emptyset
}{
    \Sigma_{final} \vdash \iota_{\text{end}} \implies \textsc{Err(DanglingPtr)}
}
\and
\inferrule*[Lab=\textbf{\textsf{E-Layout}}]{
    v \in \mathcal{V}_{out} \land \texttt{Type}(v) = \texttt{Vec}<T> \\ 
    p = v.\texttt{ptr}, \ell = v.\texttt{len}, c = v.\texttt{cap} \\
    (\sigma_{L}(p).\text{size} < c \cdot \text{sizeof}(T)) \lor 
    (\ell > c) \lor
    (\sigma_{L}(p).\text{align} < \text{align\_of}(T)) \lor
    (c \cdot \text{sizeof}(T) > \texttt{isize::MAX})
}{
    \Sigma_{final} \vdash \iota_{\text{end}} \implies \textsc{Err(InvalidLayout)}
}
\and
\inferrule*[Lab=\textbf{\textsf{E-Leak}}]{
    h \in \mathcal{HO} \\ 
    \texttt{Dropped} \notin \sigma_O(h) \\
    \forall v \in \mathcal{V}_{out}. h \notin \mathit{Own}(v) \land h \notin \mathit{Pt}(v)
}{
    \Sigma_{final} \vdash \iota_{\text{end}} \implies \textsc{Err(Leak)}
}
\end{mathpar}
\caption{Soundness Violation Detection Rules (Part 1: Boundary-Level Checks).}
\label{fig:detection-rules-1}
\end{figure*}

\subsubsection{Inter-Procedural Analysis via Full Inlining}
\label{subsubsec:inter-procedural}

Given that \texttt{unsafe} code regions in Rust typically involve shallow call
chains and static dispatch (e.g., calls to \texttt{Vec::set\_len} or intrinsic
functions), we adopt a \textit{full inlining} approach for inter-procedural
analysis.
This strategy offers full context sensitivity by performing a recursive,
depth-first traversal of the call graph.

\paragraph{Recursive traversal.}
When the analysis encounters a call \texttt{d = F(args)},
it temporarily switches to the callee and analyzes \texttt{F} in the caller context.
Concretely, we create a fresh analysis context for \texttt{F} and initialize its entry
state from the caller's current state.
We then model parameter passing by binding actual arguments to the callee's formal parameters.
After the callee reaches a fixed point, we map the abstract state of its return value
back to the destination \texttt{d} in the caller,
and resume propagation at the successor of the call site.
This on-the-fly inlining preserves the caller's context
and makes aliasing, lifecycle, and layout constraints flow naturally across calls,
without building a separate inter-procedural CFG ahead of time.

\paragraph{Recursion and external calls.}
To avoid non-termination, we impose an inline depth limit (default $k=3$).
If the limit is reached, or the callee body is unavailable (e.g., FFI),
we use a conservative summary.
The summary may mutate any reachable state of mutable arguments,
and it returns an unconstrained value ($\top$).

\begin{figure*}[h]
\footnotesize
\begin{mathpar}
\inferrule*[Lab=\textbf{\textsf{E-Access}}]{
    \iota \in \{\text{read}(p), \text{write}(p)\} \\ o \in \mathit{Pt}(p) \\ \text{sizeof}(T) > 0 \\ S = \sigma_O(o) \cap \text{InvalidSet} \neq \emptyset
}{
    {
    \Sigma \vdash \iota \implies 
    \begin{cases} 
        \textsc{Err(UAF)} & \text{if } \texttt{Dropped} \in S \\
        \textsc{Err(UBD)} & \text{if } \texttt{Uninit} \in S \\
        \textsc{Err(UAM)} & \text{if } (\texttt{Moved} \in S) \lor (\texttt{Forgotten} \in S)
    \end{cases}
    }
}
\and
\inferrule*[Lab=\textbf{\textsf{E-DoubleFree}}]{
    \iota = \texttt{drop}(x) \\ 
    (\exists h \in \mathit{Own}(x). \texttt{Dropped} \in \sigma_O(h)) \lor (\sigma_O(x) \cap \{\texttt{Moved}, \texttt{Forgotten}, \texttt{Uninit}\} \neq \emptyset)
}{
    \Sigma \vdash \iota \implies \textsc{Err(DoubleFree)}
}
\and
\inferrule*[Lab=\textbf{\textsf{E-RefCreation}}]{
    \iota = \text{mk\_ref}(p) \\ o \in \mathit{Pt}(p)
}{
    {
    \Sigma \vdash \iota \implies 
    \begin{cases}
        \textsc{Err(InvalidRef)} & \text{if } (\text{sizeof}(T) > 0) \land (\sigma_O(o) \cap \text{InvalidSet} \neq \emptyset) \\
        \textsc{Err(Null/Align)} & \text{if } \sigma_{L}(p).\text{align} < \text{align\_of}(T)
    \end{cases}
    }
}
\and
\inferrule*[Lab=\textbf{\textsf{E-OOB/MA}}]{
    \iota = \text{access}(p) \\ (\_, \text{size}, [l, u], \text{align}, \_, \_) = \sigma_{L}(p)
}{
    {
    \Sigma \vdash \iota \implies 
    \begin{cases}
        \textsc{Err(OOB)} & \text{if } l < 0 \lor (u + \text{sizeof}(T)) > \text{size} \\
        \textsc{Err(MA)} & \text{if } \text{align} < \text{align\_of}(T)
    \end{cases}
    }
}
\and
\inferrule*[Lab=\textbf{\textsf{E-NullDeref}}]{
    \iota \in \{\text{deref}(p), \text{mk\_ref}(p)\} \\
    \sigma_{L}(p).\mathit{null} = \texttt{MaybeNull}
}{
    \Sigma \vdash \iota \implies \textsc{Err(NullDeref)}
}
\and
\inferrule*[Lab=\textbf{\textsf{E-SizeMismatch}}]{
    \iota \in \{\texttt{ptr::read/write}\dots\} \\
    \sigma_{L}(p).\text{size} < \text{sizeof}(T)
}{
    \Sigma \vdash \iota \implies \textsc{Err(SizeMismatch)}
}
\and
\inferrule*[Lab=\textbf{\textsf{E-Proj}}]{
    \iota = \text{Proj}(p, f) \\ 
    \delta = \text{offset}(f) \\
    \sigma_L(p).\text{size} < \delta + \text{sizeof}(f)
}{
    \Sigma \vdash \iota \implies \textsc{Err(InvalidProj)}
}
\and
\inferrule*[Lab=\textbf{\textsf{E-PtrArith}}]{
    \iota = p.\texttt{add}(n) \\ (\_, \text{size}, [\_, u], \_, \_, \_) = \sigma_{L}(p) \\ u + n \cdot \text{sizeof}(T) > \text{size}
}{
    \Sigma \vdash \iota \implies \textsc{Err(InvalidPtrArith)}
}
\and
\inferrule*[Lab=\textbf{\textsf{E-SliceInvalid}}]{
    \iota = \texttt{slice::from\_raw}(p, len) \\ 
    (len \cdot \text{sizeof}(T) > \texttt{isize::MAX}) \lor
    (len \cdot \text{sizeof}(T) > \sigma_L(p).\text{size})
}{
    \Sigma \vdash \iota \implies \textsc{Err(InvalidSlice)}
}
\and
\inferrule*[Lab=\textbf{\textsf{E-MutateImmut}}]{
    \iota = \texttt{write}(p, \_) \\ \sigma_{L}(p).\mathit{mut} = \texttt{Immut}
}{
    \Sigma \vdash \iota \implies \textsc{Err(MutateImmutable)}
}
\end{mathpar}
\caption{Soundness Violation Detection Rules (Part 2: Instruction-Level Checks). $\text{InvalidSet} = \{\texttt{Dropped}, \texttt{Moved}, \texttt{Forgotten}, \texttt{Uninit}\}$.}
\label{fig:detection-rules-2}
\end{figure*}

\subsection{Detecting Potential Violations}
\label{subsec:violations}

We formalize detection as judgment rules of the form
$\Sigma \vdash \text{Ctx} \implies \textsc{Err}(\kappa)$.
$\text{Ctx}$ denotes either a runtime instruction $\iota$ or the function exit
point $\iota_{\text{end}}$.
The rules are summarized in Figure~\ref{fig:detection-rules-1}
and Figure~\ref{fig:detection-rules-2}.
We treat both \emph{boundary-level contract violations} and \emph{instruction-level UB}
as evidence of unsoundness.
Accordingly, we organize detection into two granularities.
The first is a boundary-level check at $\iota_{\text{end}}$,
which focuses on safety invariants exposed to safe clients.
The second is an instruction-level check during propagation,
which focuses on instruction-level UB inside unsafe code
(e.g., invalid dereference and out-of-bounds access).

\paragraph{Boundary-Level Checks}
\label{subsubsec:soundness}

The core question is whether a safe API can leak an unsafe state that enables safe callers to trigger UB.
We perform these checks at function exit ($\iota_{\text{end}}$) and apply them to
observable output variables.
We let $\mathcal{V}_{out}$ denote the set of values that escape to the caller at
exit---return values, out-parameters (e.g., mutable reference targets that remain
visible to the caller), and any other exposed references or owned values.
Our boundary checks consider only variables in $\mathcal{V}_{out}$.
\textsc{E-AliasRet} prevents double-free by ensuring no two return values claim
ownership of the \emph{same underlying resource} (checking intersection of
$\mathit{Own}$ sets).
\textsc{E-MutAlias} enforces Rust's uniqueness guarantee by detecting if multiple
return values (specifically references) alias the same location with at least
one being mutable.
\textsc{E-Dangling} ensures exposed pointers do not target invalid memory.
\textsc{E-Layout} checks that assembled containers (e.g., \texttt{Vec::from\_raw})
satisfy their validity invariants on internal components
(pointer, length, and capacity).
This includes capacity bounds, length limits, alignment, and \texttt{isize::MAX} overflow.
Finally, \textsc{E-Leak} identifies resource leaks by checking for any heap
object $h$ that remains live ($\texttt{Dropped} \notin \sigma_O(h)$) at
function exit but is unreachable from the caller (i.e., not owned or
pointed to by any variable in $\mathcal{V}_{out}$).

\paragraph{Instruction-Level Checks}
\label{subsubsec:internal}

\textsc{E-Access} guards against UAF, UBD, and UAM.
\textsc{E-DoubleFree} prevents dropping invalid or already-dropped resources.
It checks the state of the \emph{resource} rather than just the variable, which
reduces false positives from spurious aliasing.
\textsc{E-RefCreation} ensures references are created from valid, aligned,
non-null pointers.
\textsc{E-OOB/MA} and \textsc{E-NullDeref} handle standard memory errors.
\textsc{E-PtrArith} and \textsc{E-Proj} enforce in-bounds requirements for
pointer arithmetic and field projection (covering \texttt{[place-projection]}
UB).
\textsc{E-SliceInvalid} prevents the creation of invalid slices (e.g., length
exceeding allocation).

\section{Experiment Settings}
\label{sec:experiment-settings}

\subsection{Research Questions}
We evaluate \tool by answering the following research questions:
\begin{enumerate}[label=RQ\arabic*., leftmargin=*]
\item \textbf{Effectiveness}:
On a ground-truth CVE benchmark, can our framework outperform state-of-the-art tools in detecting Rust-specific UBs?
\item \textbf{Precision and Recall Analysis}:
What precision/recall trade-offs does our framework exhibit,
and what are the main sources of false positives and false negatives?
\item \textbf{Efficiency}:
Is our framework efficient enough for real-world use,
with acceptable analysis time and memory overhead on large Rust crates?
\item \textbf{Large-Scale Discovery of Unknown Bugs}:
Can our framework discover previously unreported bugs in real-world Rust crates at scale?
\end{enumerate}

\subsection{Dataset Construction}
We constructed two distinct datasets to support the studies in Section~\ref{sec:results}:

\smallskip
\noindent
\textbf{Dataset A (Known Vulnerabilities)}:
We selected 46 CVEs from the RustSec Advisory Database.
To construct a focused benchmark, we specifically targeted soundness issues caused by unsafe memory operations (e.g., buffer overflow, use-after-free, uninitialized memory) that fall within our sequential analysis scope.
Accordingly, we excluded vulnerabilities that rely on external semantics (e.g., FFI, platform-specific hardware) or application-specific logic errors (e.g., index desynchronization or complex state machine bugs).
Because many baseline tools rely on older compiler versions that often fail to compile the original crates,
we extracted the key vulnerable code into minimized libraries that faithfully reproduce the original bugs.
We manually verified these test cases and inspected the tool warnings
to ensure that any potential new bugs in the constructed code are correctly identified,
avoiding incorrect false positive reports.
We use RustSec-derived cases rather than Miri or Rudra benchmark suites as the
primary benchmark because Miri is an execution validator rather than a
static-analysis benchmark, and Rudra includes bug classes outside our current
modeled scope; all baselines are therefore compared on the same RustSec-derived
subjects.
This dataset serves as the ground truth for measuring detection effectiveness.

\smallskip
\noindent
\textbf{Dataset B (Wild Crates)}:
To evaluate the bug detection capability of \tool in the wild, we scanned the entire crates.io registry.
Due to the large volume of results, it is impractical to manually audit all of them.
Instead, we manually inspected the reports from approximately 10,000 crates to verify the findings.
From this inspected subset, we identified 83 distinct crates containing potential bugs, which form Dataset B.
These crates span diverse domains, including data structures, memory management, and no-std libraries,
with a total of over 30 million downloads.

\subsection{Baseline Selection}
We selected three representative baselines covering distinct Rust static analysis paradigms,
ensuring fair comparison against state-of-the-art tools.
All baselines used their latest stable versions and default configurations to match real-world usage.
All experiments were conducted on the same machine with an Intel Core i7 processor and 64GB of RAM, running Ubuntu 22.04 LTS.

\begin{enumerate}
    \item \textbf{MirChecker} \cite{MIRChecker2021}: MIR-level static analyzer, uses numerical/symbolic analysis for generic errors (e.g., integer overflow) but lacks Rust-specific ownership tracking.
    \item \textbf{Rudra} \cite{rudra2021}: Unsafe code pattern matcher, flags risky unsafe constructs (e.g., `ptr::copy` in unsafe blocks) but lacks semantic understanding.
    \item \textbf{SafeDrop} \cite{SafeDrop2023}: Deallocation-focused analyzer, detects double free/UAF but ignores alignment/OOB and nested ownership.
\end{enumerate}

\subsection{Implementation}
\label{sec:implementation}
Our prototype, \tool, is implemented as a Rust compiler plugin based on the MIR (Mid-level Intermediate Representation).
It uses the \texttt{rustc\_middle} and \texttt{rustc\_mir\_dataflow} libraries for MIR traversal and dataflow analysis.
For symbolic modeling of memory layouts and values, we drew inspiration from the design of MIRAI~\cite{mirai}, an abstract interpreter for Rust.
While MIRAI focuses on generic property verification (e.g., panics and standard assertions) using a heavy-weight solver-based approach,
our implementation specializes its symbolic memory model to efficiently track Rust's unique ownership and lifecycle semantics alongside spatial constraints.
For numerical analysis, we integrate the \textbf{Apron} numerical abstract domain library.
Constraint solving is handled by the \textbf{Z3} SMT solver.

\section{Results and Analysis}
\label{sec:results}

\subsection{RQ1: Effectiveness}
\label{sec:rq1}

To evaluate the effectiveness of \tool in detecting real-world vulnerabilities,
we tested it against Dataset A, which comprises 46 known CVEs.
In contrast to RQ4, which targets open-world bug discovery in the wild (Dataset B),
RQ1 evaluates detection effectiveness on a closed benchmark with ground truth
and compares against state-of-the-art baselines.
We compared our results with three baseline tools: MirChecker, Rudra, and SafeDrop.
Table \ref{tab:rq1-merged} presents the comparative results.
We count a CVE as detected if a tool reports at least one true positive alert that matches the vulnerability,
based on our manual audit of the tool outputs.

\begin{table}[htbp]
\centering
\caption{Comparison of bug detection effectiveness on 46 known vulnerabilities (Dataset A). \textbf{Detected by Type} counts the number of \textit{unique bugs} (ground-truth patterns) successfully detected by each tool for each category. \textbf{Total Bugs} is the number of detected bugs out of the 53 total ground-truth bugs. \textbf{Total CVE} is the number of unique CVEs detected out of 46. The \textbf{Recall (Bug)} column reports bug-level recall. \textbf{Total Alerts} shows the total number of warnings generated by the tool, \textbf{TP Alerts} is the number of true positive alerts, and \textbf{Precision} is computed as \textbf{TP Alerts}/\textbf{Total Alerts}. The number in parentheses in the header row indicates the number of ground-truth bugs for that specific unsoundness category.}
\label{tab:rq1-merged}
\resizebox{\linewidth}{!}{
\begin{tabular}{l c c c c c c c c c}
\toprule
\textbf{Tool} & \multicolumn{3}{c}{\textbf{Detected by Type (Bugs)}} & \textbf{Total} & \textbf{Total} & \textbf{Recall} & \textbf{TP} & \textbf{Total} & \textbf{Precision} \\
\cmidrule(lr){2-4}
 & \textbf{Uninit (24)} &  \textbf{OOB (25)} & \textbf{Own (4)} & \textbf{Bugs (53)} & \textbf{CVE (46)} & \textbf{(Bug)} & \textbf{Alerts} & \textbf{Alerts} & \\
\midrule
MirChecker & 0 & 0 & 0 & 0 & 0 & 0.0\% & 0 & 45 & 0.0\% \\
SafeDrop & 0 & 0 & 1 & 1 & 1 & 1.9\% & 1 & 4 & 25.0\% \\
Rudra & 11 & 0 & 0 & 11 & 11 & 20.9\% & 11 & 11 & 100.0\% \\
\textbf{\tool (Ours)} & \textbf{15} & \textbf{19} & \textbf{2} & \textbf{36} & \textbf{32} & \textbf{67.9\%} & \textbf{64} & 124 & \textbf{51.6\%} \\
\bottomrule
\end{tabular}
}
\end{table}

\tool detected 32 out of 46 vulnerabilities (69.6\%),
significantly outperforming the best baseline, Rudra,
which detected only 11 (23.9\%).
As shown in Table \ref{tab:rq1-merged},
\tool demonstrates superior effectiveness across all vulnerability categories.
At the bug level (ground-truth patterns), \tool successfully covered \textbf{36 out of 53 bugs (67.9\%)},
including 15 uninitialized memory bugs and 19 out-of-bounds bugs,
whereas Rudra only covered 11 bugs in total.
Compared with Rudra, \tool identified an additional 21 CVEs.
The 53 ground-truth bug patterns further refine into six categories:
11 uninitialized-memory bugs, 13 length or state-consistency bugs,
25 out-of-bounds or pointer-boundary bugs, 1 use-after-free bug,
1 double-ownership bug, and 2 dangling-pointer bugs.
Table \ref{tab:rq1-merged} aggregates these fine-grained labels into
three analysis-facing groups:
uninitialized-memory and length/state-consistency bugs are reported under
uninitialized memory, out-of-bounds and pointer-boundary bugs under
out-of-bounds, and use-after-free, double ownership, and dangling pointers
under dangling pointers.

\begin{figure*}[t]
\begin{minipage}[t]{0.47\textwidth}
\begin{minipage}[c][0.17\textheight][c]{\linewidth}
\input{code/listing-05.tex}
\end{minipage}
\captionsetup{skip=2pt}
\captionof{listing}{Buggy code for \texttt{RUSTSEC-2023-0078} in \texttt{tracing}: take field pointers and then call \texttt{mem::forget}.}
\label{lst:tracing}
\end{minipage}\hfill
\begin{minipage}[t]{0.47\textwidth}
\begin{minipage}[c][0.17\textheight][c]{\linewidth}
\input{code/listing-06.tex}
\end{minipage}
\captionsetup{skip=2pt}
\captionof{listing}{An instance of \texttt{uninitialized} used for entropy generation.}
\label{lst:temporary-uninit}
\end{minipage}
\end{figure*}

To further illustrate \tool's detection capability,
we highlight three representative cases missed by all baselines.
One example is \texttt{RUSTSEC-2023-0078} in the well-known \texttt{tracing} crate maintained by \texttt{tokio-rs}.
Listing~\ref{lst:tracing} shows the vulnerable pattern: it takes pointers to fields and then calls \texttt{mem::forget(self)}.
This is UB because \texttt{mem::forget(self)} consumes (moves) \texttt{self}; after that point,
\texttt{self} is no longer live and the compiler may reuse its stack storage,
invalidating the saved pointers to its fields.
The advisory was fixed by replacing \texttt{mem::forget} with \texttt{ManuallyDrop}
in \texttt{Instrumented::into\_inner}~\footnote{\url{https://github.com/tokio-rs/tracing/pull/2765}},
which keeps the storage valid when taking field pointers.
In our model, \texttt{mem::forget(self)} transitions \texttt{self} to the \texttt{Forgotten} state,
which represents a semantic move with destructor suppression.
Subsequent reads through pointers derived from \texttt{self} are then reported as a use-after-move alarm.
Reading from uninitialized memory is undefined behavior in Rust.
In \texttt{RUSTSEC-2018-0022}, the developer used \texttt{std::mem::uninitialized} to read stack garbage for entropy.
For example, Listing~\ref{lst:temporary-uninit} shows an instance reported in the \texttt{temporary} crate~\footnote{\url{https://github.com/stainless-steel/temporary/issues/2}}.
\tool's lifecycle domain tracks the \texttt{Uninit} state of the returned value and reports the subsequent read as an uninitialized-memory violation.
For pointer bounds, we detect an unsoundness of \texttt{wrflib}~\footnote{\url{https://rustsec.org/advisories/RUSTSEC-2025-0072.html}} in \texttt{RUSTSEC-2025-0072}.
The functions under \texttt{wrflib::byte\_extract} are wrappers over pointer offset (e.g., \texttt{data.as\_ptr().add(offset)}) without sufficient bounds checks.
Our layout domain models the spatial attributes of the pointer and the bounds of the allocation.
By propagating these constraints to the dereference point,
it can determine that \texttt{offset} may exceed the object's size, pinpointing the out-of-bounds violation.

Baselines missed these cases because they require semantic reasoning over object states and derived pointers,
rather than matching a specific risky API signature.
\tool detected more than 20 vulnerabilities that were missed by all other baselines.
This demonstrates that \tool fills a significant gap in Rust static analysis.

\subsection{RQ2: Precision and Recall Analysis}
\label{sec:rq2}

While RQ1 focuses on effectiveness (recall),
this section analyzes the trade-off between precision and recall by inspecting false positives and false negatives.
As shown in Table \ref{tab:rq1-merged}, \tool generated a total of 124 alerts.
Among them, 64 are true positive alerts (which successfully cover the 36 ground-truth bugs)
and 60 are false positive alerts, resulting in an alert-level precision of 51.6\%.
Despite its high recall compared to baselines,
\tool missed 14 vulnerabilities (30.4\%) and 17 ground-truth bugs (32.1\%).
We organize the main sources of imprecision by domain (Section~\ref{sec:approach}).

\subsubsection{False Positives}
We analyzed the 60 false positives to identify the primary sources of imprecision.
The largest share (51.7\%, 31/60) stems from the \textbf{layout domain},
primarily because our layout facts are conservative lower bounds
and our arithmetic reasoning uses widening and interval abstractions.
When indices depend on complex non-linear operations (e.g., bitwise arithmetic) or external inputs,
our solver may fail to derive tight constraints.
In such cases, we keep conservative offset intervals,
which can lead to potential out-of-bounds or misalignment alarms.
The \textbf{lifecycle domain} contributes another 35.0\% (21/60),
often caused by control-flow merges and opaque calls.
Our lifecycle domain represents uncertainty as sets of states (e.g., \{\texttt{Live}, \texttt{Uninit}\}),
so conditional initialization patterns naturally lead to ``maybe uninitialized'' alarms.
This is amplified at external call boundaries (e.g., FFI) and at recursion cutoffs,
where we conservatively over-approximate effects and may lose the evidence needed to prove initialization.
Finally, the \textbf{ownership domain} accounts for 13.3\% (8/60),
typically arising from path-insensitive aliasing approximations in functions with complex control flows,
where the tool may incorrectly infer double frees by merging disjoint paths.

Dataset B should not be interpreted as a closed-world precision experiment,
because the remaining warnings were not exhaustively labeled.
Instead, it measures alarm density, triage workload, and validated discovery yield.
Among the approximately 10,000 manually inspected crates, \tool flagged 378 crates
and produced 2,639 raw warnings, which were de-duplicated into 1,736 unique warnings
during triage.
This bounded workload led to 114 Miri-validated bugs across 83 crates;
45 have been acknowledged by developers or RustSec experts, and 27 have been fixed.
For alert-volume context, Rudra successfully compiled and analyzed 2,892 crates
in the same 10,000-crate set, flagged 119 crates, and produced 326 warnings.
On the same Rudra-compilable subset, \tool flagged 141 crates and produced
1,296 raw warnings, deduplicated to 784 unique warnings.
These numbers describe workload and triage scale, not closed-world precision.

\subsubsection{False Negatives}
We manually inspected the 14 missed vulnerabilities and grouped their primary causes
into four scope or precision boundaries.
First, 7 missed CVEs require library or API semantics not explicitly modeled by the
current implementation, such as \texttt{Cell}, alignment-related allocation APIs,
\texttt{debug\_assert}, or specialized crate APIs.
For example, \tool missed \texttt{RUSTSEC-2023-0017} (maligned) because we do not yet model \texttt{Layout::from\_size\_align},
leading to missing layout constraints for the allocated pointer.
Second, 3 require more precise numeric, loop, or iterator reasoning than our current
interval abstraction maintains.
Third, 2 require per-element array or initialization-state tracking finer than our
current memory abstraction.
Finally, 2 depend on external or cross-boundary logic state, such as cache behavior
or raw-pointer state changes across method sequences.
Despite these misses, \tool's detection capability remains significantly higher than all baselines.
The missed cases identify engineering and modeling boundaries rather than invalidating
the modeled warning rules.

\subsection{RQ3: Efficiency}
We evaluated the runtime efficiency of \tool to determine its suitability for real-world development workflows.
We report both end-to-end running time and the split between UnsafeChecker's analysis pass
and compilation/non-analysis overhead.
The end-to-end number reflects user-visible cost, while the split isolates the cost
of the analysis itself.
We measured the running time and memory overhead for 9 representative crates from Dataset B,
ranging from small utilities to large-scale applications.

Table \ref{tab:rq3-efficiency} details the performance metrics for these crates.
Across the 9 crates, the median end-to-end running time is 11.4s,
with a maximum of 264.0s.
For small libraries ($<$10k LOC), the analysis typically completes within 10 seconds.
For large crates ($>$100k LOC), the analysis finishes within 5 minutes,
and the slowest case we measured (\texttt{meilisearch}) completes in about 264s.
The median memory usage is 475.2 MB and the peak is 4.3 GB,
which fits typical developer machines.

\begin{table}[htbp]
\centering
\scriptsize
\caption{Efficiency of \tool on Real-World Crates.
\textbf{A(s)} reports UnsafeChecker analysis time, \textbf{C/O(s)} reports compilation and other non-analysis overhead, and \textbf{T(s)} reports end-to-end time.}
\label{tab:rq3-efficiency}
\resizebox{\linewidth}{!}{
\begin{tabular}{lrrrrr|lrrrrr|lrrrrr}
\toprule
\multicolumn{6}{c|}{\textbf{Small ($<$10k LOC)}} & \multicolumn{6}{c|}{\textbf{Medium (10k-100k LOC)}} & \multicolumn{6}{c}{\textbf{Large ($>$100k LOC)}} \\
\cmidrule(r){1-6} \cmidrule(lr){7-12} \cmidrule(l){13-18}
\textbf{Crate} & \textbf{LOC} & \textbf{A(s)} & \textbf{C/O(s)} & \textbf{T(s)} & \textbf{M(MB)} & \textbf{Crate} & \textbf{LOC} & \textbf{A(s)} & \textbf{C/O(s)} & \textbf{T(s)} & \textbf{M(MB)} & \textbf{Crate} & \textbf{LOC} & \textbf{A(s)} & \textbf{C/O(s)} & \textbf{T(s)} & \textbf{M(MB)} \\
\midrule
bitflags & 5.0k & 0.6 & 0.3 & 0.9 & 118.7 & serde & 33.3k & 4.4 & 1.7 & 6.1 & 275.7 & diesel & 126.8k & 89.2 & 24.3 & 113.5 & 1350.2 \\
log & 3.8k & 1.1 & 0.5 & 1.6 & 109.0 & regex & 96.5k & 6.8 & 1.8 & 8.6 & 505.2 & meilisearch & 244.6k & 211.7 & 52.3 & 264.0 & 4329.2 \\
bytes & 7.1k & 8.7 & 2.7 & 11.4 & 170.2 & tokio & 93.1k & 14.6 & 4.5 & 19.1 & 475.2 & nushell & 287.6k & 70.9 & 23.7 & 94.6 & 2853.6 \\
\bottomrule
\end{tabular}
}
\end{table}

The running time does not strictly correlate with LOC,
as it also depends on code complexity (e.g., heavy use of macros or complex control flows).
For instance, \texttt{diesel} (126.8k LOC) takes longer (113.5s) than \texttt{nushell} (287.6k LOC, 94.6s).
Overall, these results show that our shared-state analysis is efficient enough for daily use.
At ecosystem scale, running on a compute server, \tool analyzed over 100,000 crates within two days.

We also compared the efficiency of \tool against the baselines on Dataset A (the CVE benchmark).
The reported times include both Rust compilation and tool execution.
\tool analyzed all 46 CVE cases in 35.7 seconds (0.78s per case on average).
In contrast, Rudra took 179.7 seconds ($5\times$ slower), and MirChecker required 781.9 seconds ($>20\times$ slower).
SafeDrop achieved a similar speed (36.1s), but as shown in RQ1, it only detected one vulnerability in this dataset.
This indicates that \tool keeps the analysis cost practical while preserving the RQ1 effectiveness advantage.

\subsection{RQ4: Large-Scale Discovery of Unknown Bugs}
We applied \tool to the entire crates.io registry to demonstrate its scalability.
Running on a compute server, \tool analyzed over 100,000 crates within two days.
Unlike RQ1--RQ3, which evaluate against a ground-truth CVE benchmark,
RQ4 targets open-world bug discovery, where no complete ground truth exists.
Due to the high volume of alarms from over 100,000 crates,
we employed random sampling to select approximately 10,000 crates for manual verification.
This process yielded 83 distinct crates with potential bugs (Dataset B).
Our verification criterion is strict:
a warning is considered a bug only if we can construct a Proof-of-Concept (PoC) triggering undefined behavior (UB) on MIRI via safe public APIs.

\begin{figure*}[t]
    \centering
    \captionsetup[subfigure]{skip=1pt}
    \captionsetup{skip=2pt}
    \begin{subfigure}{0.32\linewidth}
        \centering
        \includegraphics[width=0.9\linewidth]{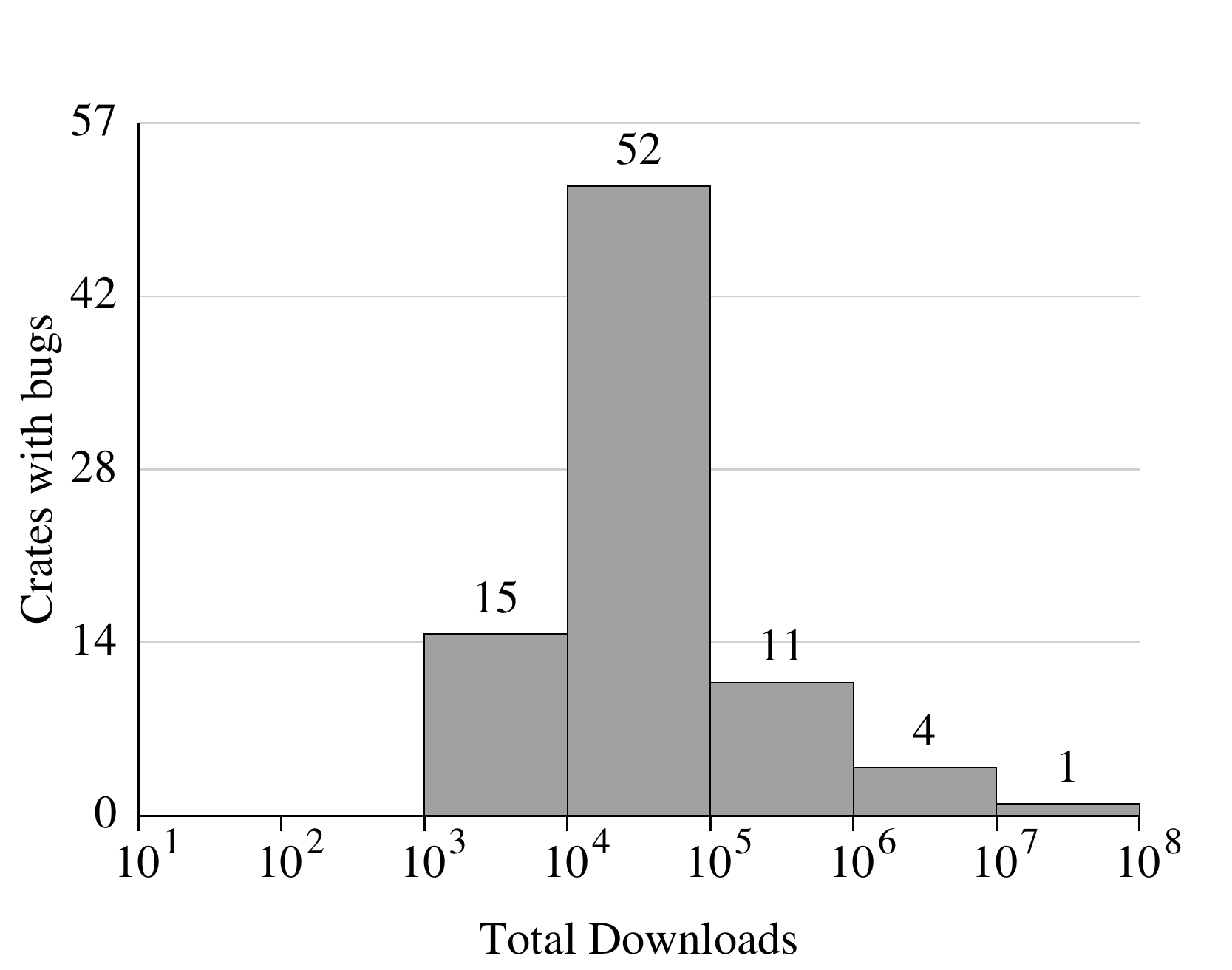}
        \caption{Total Downloads}
        \label{fig:rq4-downloads-total}
    \end{subfigure}
    \hfill
    \begin{subfigure}{0.32\linewidth}
        \centering
        \includegraphics[width=0.9\linewidth]{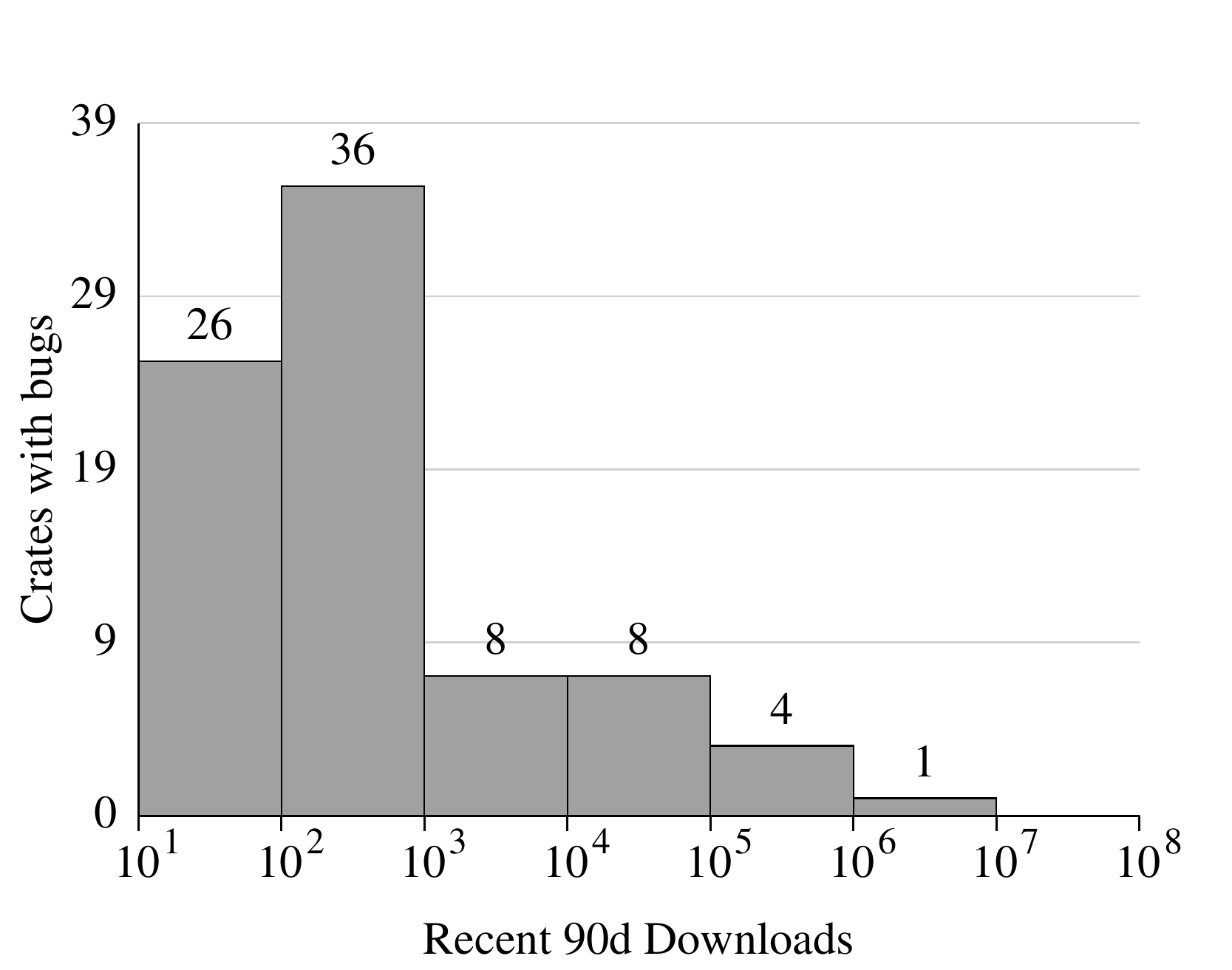}
        \caption{Recent 90-day Downloads}
        \label{fig:rq4-downloads-recent}
    \end{subfigure}
    \hfill
    \begin{subfigure}{0.32\linewidth}
        \centering
        \includegraphics[width=0.9\linewidth]{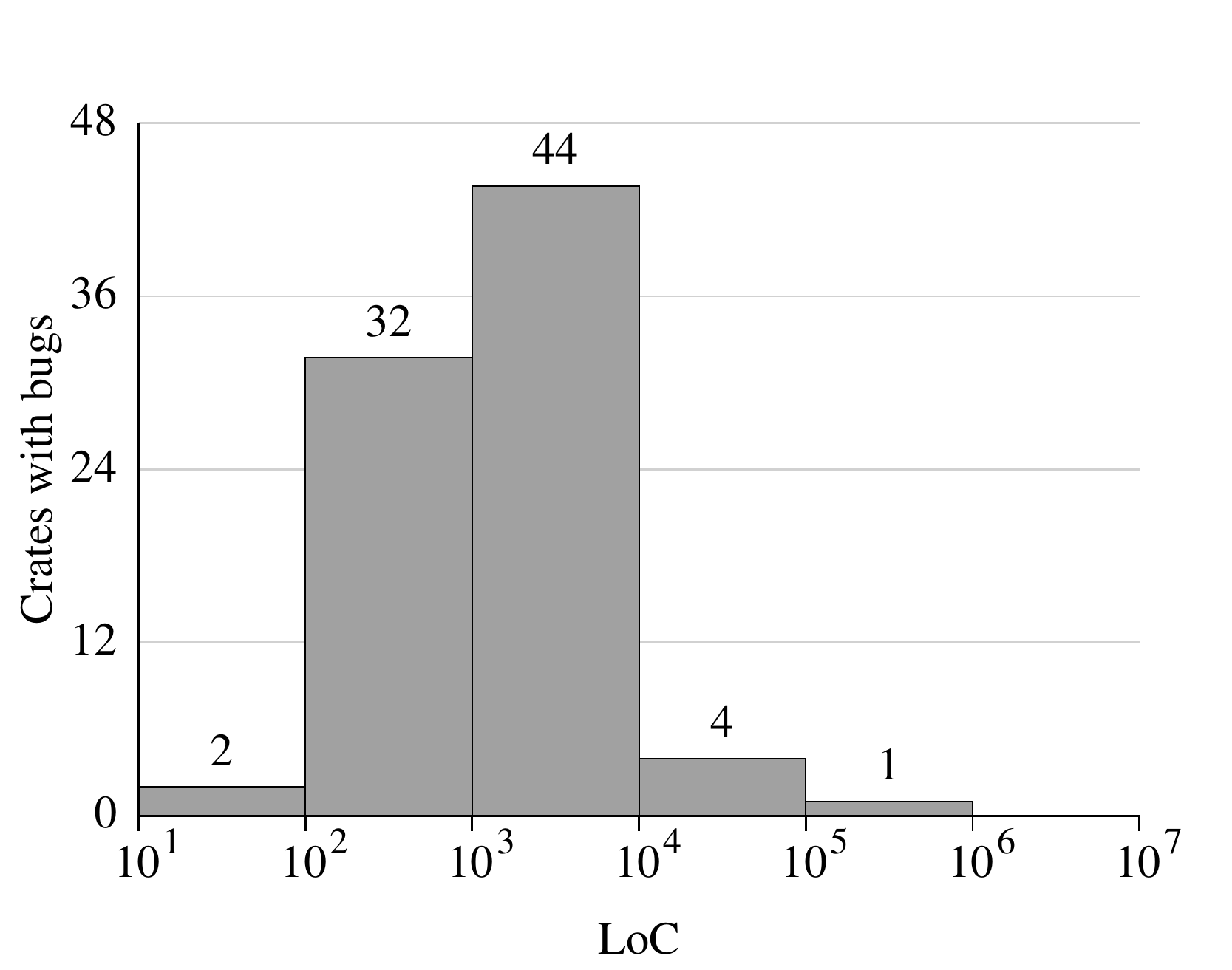}
        \caption{LoC Distribution}
        \label{fig:rq4-loc}
    \end{subfigure}
    \caption{Characteristics of 83 Flagged Crates (114 Miri-Validated Bugs)}
    \label{fig:rq4-stats}
\end{figure*}

The results are summarized in Table \ref{tab:rq4-bugs}. Within the manually inspected subset,
\tool identified \textbf{114 Miri-validated bugs} across 83 crates.
Importantly, we manually validated these reports by constructing executable PoCs that trigger UB in Miri through safe public APIs.
The PoCs were manually written; \tool's reports guided their construction by
identifying the unsafe operation, the violated invariant, and the relevant state trace.
This result should be read as open-world discovery yield rather than alert-level precision.
We reported all these findings to the crate maintainers.
As of the time of writing, developers or RustSec experts have acknowledged \textbf{45 bugs} and fixed \textbf{27 of them}.
Eleven validated findings have received public vulnerability identifiers.
The remaining reports are currently pending developer response.
The flagged crates range from small utilities to large libraries, confirming the tool's scalability.
Figure \ref{fig:rq4-stats} details the characteristics of the flagged crates.
These libraries are widely used and actively maintained, as reflected by their download statistics,
meaning that they affect the Rust ecosystem's security.
Vulnerabilities span 34 categories, with \textit{data-structures}, \textit{no-std}, and \textit{memory-management} being most frequent,
indicating that even foundational libraries often fail to guarantee memory safety.

The issues detected by \tool encompass a broad spectrum of memory safety
violations, along with a small set of auxiliary checks.
They are not concentrated in a fixed \texttt{ptr::read} or \texttt{mem::forget}
pattern:
at the bug level, these two APIs account for only 5 and 1 of the 53 Dataset A
bug patterns covered by \tool, and 19 and 4 of the 114 Dataset B bugs,
respectively.
In Dataset B, the validated bugs involve 129 related unsafe-API call sites
across 21 APIs, including pointer arithmetic and offset APIs, unchecked
slice/string construction APIs, raw ownership or allocation-takeover APIs, and
type or initialization invariant bypasses.
In Table \ref{tab:rq4-bugs}, we categorize these 114 reports by their observable
manifestations---such as out-of-bounds (OOB) access or use-after-free (UAF)---to
facilitate reporting and comparison.
However, \tool detects these vulnerabilities by identifying invariant violations over the facts maintained in its shared abstract state.
Specifically, violations in the \textit{layout domain} account for the majority of cases, including \textbf{out-of-bounds} (60 cases), \textbf{misaligned access} (11 cases), and various forms of \textbf{invalid pointers} (e.g., null dereferences).
These issues typically arise from incorrect pointer arithmetic or unchecked casts that violate spatial memory constraints.
The \textit{ownership domain} and \textit{lifecycle domain} are critical for detecting temporal safety issues; they identified \textbf{ownership duplication} (17 cases) and \textbf{dangling pointers} (7 cases) by tracking invalid object state transitions, such as accessing moved values or dropping the same resource twice.
Beyond these, \tool also flagged a small set of auxiliary issues, including \textbf{resource leaks} (5 cases),
\textbf{integer overflow}, \textbf{division by zero}, and \textbf{UTF-8 validation}.
These issues are not strictly memory-unsafe by themselves (e.g., memory leaks are safe in Rust), but they frequently indicate logical errors or appear in
index and size computations that feed into unsafe memory accesses.

\begin{table}[htbp]
\centering
\footnotesize
\setlength{\tabcolsep}{3pt}
\renewcommand{\arraystretch}{0.95}
\caption{
Summary of detected, confirmed, and fixed issues by type.
Abbreviations:
OOB (Out-of-Bounds),
OD (Ownership Duplication),
MA (Misaligned Access),
IO (Integer Overflow),
IP (Invalid Pointer),
DP (Dangling Pointer),
DBZ (Division by Zero),
UTF8 (UTF-8 Validation),
RL (Resource Leak),
NP (Null Pointer),
UM (Uninitialized Memory).
\textbf{Crates} counts unique crates per type (a crate may appear in multiple columns), so summing this row will over-count and can exceed the total.
}
\label{tab:rq4-bugs}
\begin{tabular}{l c c c c c c c c c c c c c}
\toprule
 & OOB & OD & MA & IO & IP & DP & DBZ & UTF8 & RL & NP & UM & Other & Total \\
\midrule
\textbf{Crates} & 42 & 13 & 8 & 3 & 3 & 6 & 1 & 1 & 5 & 2 & 1 & 2 & \textbf{83} \\
\textbf{Detected} & 60 & 17 & 11 & 3 & 3 & 7 & 1 & 1 & 5 & 2 & 1 & 3 & \textbf{114} \\
\textbf{Confirmed} & 20 & 10 & 4 & 2 & 2 & 2 & 1 & 1 & 1 & 1 & 1 & 0 & \textbf{45} \\
\textbf{Fixed} & 13 & 6 & 0 & 1 & 1 & 1 & 1 & 1 & 1 & 1 & 1 & 0 & \textbf{27} \\
\bottomrule
\end{tabular}
\end{table}

Table~\ref{tab:rq4-bugs} shows that layout-related issues dominate the reports: Out-of-Bounds accounts for 52.6\% (60/114) and misaligned access accounts for 9.6\% (11/114).
Temporal and ownership-related issues are also common, including ownership duplication (17/114) and dangling pointers (7/114).
These distributions align with our three-component classification, with most confirmed bugs rooted in violations of layout, object-validity, or ownership invariants.

\subsubsection{Case Studies}
We present four case studies of real-world bugs. Cases 1, 3, and 4 illustrate how our framework detects composite UBs missed by baselines, while Case 2 demonstrates an auxiliary check for resource leaks.

\textbf{Case 1: Dangling Pointer in \texttt{multiversx\_chain\_vm}} (132k downloads).
We detected a critical drop-safety violation in \texttt{multiversx\_chain\_vm}
that leads to a dangling pointer.
The function \texttt{with\_shared\_mut\_ref}, shown in Listing \ref{lst:multiversx-bug}, uses \texttt{unsafe} code to
\textbf{duplicate ownership} of a value behind a mutable reference (\texttt{\&mut T})
using \texttt{std::ptr::read}.
This operation performs a bitwise copy of the underlying value without consuming the original,
effectively creating two owners for the same resource: one in the temporary \texttt{Rc} and one still behind the reference.
If the user-provided closure \texttt{f} panics,
the temporary \texttt{Rc} is dropped during stack unwinding,
deallocating the underlying resource \texttt{obj}.
However, the original slot still contains a stale copy of \texttt{obj}.
As a result, the caller can later access freed memory through \texttt{t},
and dropping \texttt{t} triggers a double free.

\begin{listing}[htbp]
\begin{Verbatim}[commandchars=\\\{\},numbersep=2pt, fontsize=\scriptsize, numbers=left]
\PYG{k}{pub}\PYG{+w}{ }\PYG{k}{fn}\PYG{+w}{ }\PYG{n+nf}{with\PYGZus{}shared\PYGZus{}mut\PYGZus{}ref}\PYG{o}{\PYGZlt{}}\PYG{n}{T}\PYG{p}{,}\PYG{+w}{ }\PYG{n}{F}\PYG{o}{\PYGZgt{}}\PYG{p}{(}\PYG{n}{t}\PYG{p}{:}\PYG{+w}{ }\PYG{k+kp}{\PYGZam{}}\PYG{n+nc}{mut}\PYG{+w}{ }\PYG{n}{T}\PYG{p}{,}\PYG{+w}{ }\PYG{n}{f}\PYG{p}{:}\PYG{+w}{ }\PYG{n+nc}{F}\PYG{p}{)}\PYG{+w}{ }\PYG{p}{\PYGZob{}}
\PYG{+w}{    }\PYG{k}{unsafe}\PYG{+w}{ }\PYG{p}{\PYGZob{}}
\PYG{+w}{        }\PYG{c+c1}{// ... (extract obj) ...}
\PYG{+w}{        }\PYG{k+kd}{let}\PYG{+w}{ }\PYG{n}{obj}\PYG{+w}{ }\PYG{o}{=}\PYG{+w}{ }\PYG{n}{std}\PYG{p}{::}\PYG{n}{ptr}\PYG{p}{::}\PYG{n}{read}\PYG{p}{(}\PYG{n}{t}\PYG{p}{)}\PYG{p}{;}
\PYG{+w}{        }\PYG{k+kd}{let}\PYG{+w}{ }\PYG{n}{obj\PYGZus{}rc}\PYG{+w}{ }\PYG{o}{=}\PYG{+w}{ }\PYG{n}{Rc}\PYG{p}{::}\PYG{n}{new}\PYG{p}{(}\PYG{n}{obj}\PYG{p}{)}\PYG{p}{;}
\PYG{+w}{        }\PYG{n}{f}\PYG{p}{(}\PYG{n}{obj\PYGZus{}rc}\PYG{p}{.}\PYG{n}{clone}\PYG{p}{(}\PYG{p}{)}\PYG{p}{)}\PYG{p}{;}\PYG{+w}{       }\PYG{c+c1}{// [2] Panic drops obj\PYGZus{}rc (Free \PYGZsh{}1); skips [3]}
\PYG{+w}{        }\PYG{n}{std}\PYG{p}{::}\PYG{n}{ptr}\PYG{p}{::}\PYG{n}{write}\PYG{p}{(}\PYG{n}{t}\PYG{p}{,}\PYG{+w}{ }\PYG{o}{..}\PYG{p}{.}\PYG{p}{)}\PYG{p}{;}\PYG{+w}{ }\PYG{c+c1}{// [3] Restore t (Only on normal return)}
\PYG{+w}{    }\PYG{p}{\PYGZcb{}}
\PYG{+w}{    }\PYG{c+c1}{// ...}
\PYG{p}{\PYGZcb{}}
\end{Verbatim}

\caption{Double Free vulnerability in \texttt{multiversx\_chain\_vm} triggered by panic unwinding.}
\label{lst:multiversx-bug}
\end{listing}

\textbf{Case 2: Resource Leak in \texttt{bytes-kman}}.
While memory leaks are strictly considered safe in Rust's semantic model (e.g., \texttt{Box::leak} is a safe function), they often indicate logical bugs in resource management.
As an auxiliary check, \tool identifies such resource leaks.
We detected a significant leak in the \texttt{from\_bytes} implementation for arrays in \texttt{bytes-kman}.
As shown in Listing \ref{lst:bytes-kman-bug}, the code first collects deserialized elements into a temporary \texttt{Vec<T>}.
To return a fixed-size array \texttt{[T; LEN]}, it converts the vector into a boxed slice and calls \texttt{Box::leak} to obtain a static reference.
Finally, it uses \texttt{ptr::read} to copy the data from this static reference into the return value.
Although \texttt{Box::leak} intentionally extends the lifetime of the heap memory, \texttt{ptr::read} only performs a bitwise copy without taking ownership of the backing allocation.
Consequently, the leaked heap memory becomes unreachable and is never deallocated.
\tool flagged this issue by checking that the heap object remains live at function exit but becomes unreachable from the caller (Rule \textsf{E-Leak}).

\begin{listing}[htbp]
\begin{Verbatim}[commandchars=\\\{\},numbersep=2pt, fontsize=\scriptsize, numbers=left]
\PYG{k}{fn}\PYG{+w}{ }\PYG{n+nf}{from\PYGZus{}bytes}\PYG{p}{(}\PYG{n}{buffer}\PYG{p}{:}\PYG{+w}{ }\PYG{k+kp}{\PYGZam{}}\PYG{n+nc}{mut}\PYG{+w}{ }\PYG{n}{TBuffer}\PYG{p}{)}\PYG{+w}{ }\PYG{p}{\PYGZhy{}\PYGZgt{}}\PYG{+w}{ }\PYG{n+nb}{Option}\PYG{o}{\PYGZlt{}}\PYG{n+nb+bp}{Self}\PYG{o}{\PYGZgt{}}\PYG{+w}{ }\PYG{p}{\PYGZob{}}
\PYG{+w}{    }\PYG{k+kd}{let}\PYG{+w}{ }\PYG{k}{mut}\PYG{+w}{ }\PYG{n}{res}\PYG{+w}{ }\PYG{o}{=}\PYG{+w}{ }\PYG{n+nb}{Vec}\PYG{p}{::}\PYG{n}{with\PYGZus{}capacity}\PYG{p}{(}\PYG{n}{LEN}\PYG{p}{)}\PYG{p}{;}
\PYG{+w}{    }\PYG{c+c1}{// ... (fill res) ...}
\PYG{+w}{    }\PYG{k+kd}{let}\PYG{+w}{ }\PYG{n}{res}\PYG{+w}{ }\PYG{o}{=}\PYG{+w}{ }\PYG{n+nb}{Box}\PYG{p}{::}\PYG{n}{leak}\PYG{p}{(}\PYG{n}{res}\PYG{p}{.}\PYG{n}{into\PYGZus{}boxed\PYGZus{}slice}\PYG{p}{(}\PYG{p}{)}\PYG{p}{)}\PYG{p}{;}\PYG{+w}{ }\PYG{c+c1}{// [1] Leak heap memory}
\PYG{+w}{    }\PYG{k+kd}{let}\PYG{+w}{ }\PYG{n}{res}\PYG{+w}{ }\PYG{o}{=}\PYG{+w}{ }\PYG{k}{unsafe}\PYG{+w}{ }\PYG{p}{\PYGZob{}}\PYG{+w}{ }
\PYG{+w}{        }\PYG{n}{std}\PYG{p}{::}\PYG{n}{ptr}\PYG{p}{::}\PYG{n}{read}\PYG{p}{(}\PYG{n}{res}\PYG{p}{.}\PYG{n}{as\PYGZus{}ptr}\PYG{p}{(}\PYG{p}{)}\PYG{+w}{ }\PYG{k}{as}\PYG{+w}{ }\PYG{o}{*}\PYG{k}{const}\PYG{+w}{ }\PYG{p}{[}\PYG{n}{T}\PYG{p}{;}\PYG{+w}{ }\PYG{n}{LEN}\PYG{p}{]}\PYG{p}{)}\PYG{+w}{ }\PYG{c+c1}{// [2] Copy bits; [1] never freed!}
\PYG{+w}{    }\PYG{p}{\PYGZcb{}}\PYG{p}{;}
\PYG{+w}{    }\PYG{n+nb}{Some}\PYG{p}{(}\PYG{n}{res}\PYG{p}{)}
\PYG{p}{\PYGZcb{}}
\end{Verbatim}

\caption{Memory Leak in \texttt{bytes-kman} due to \texttt{Box::leak} and \texttt{ptr::read}.}
\label{lst:bytes-kman-bug}
\end{listing}

\textbf{Case 3: Ownership Duplication in \texttt{doubly}}.
We identified a critical \textbf{double free} in the \texttt{pop\_back} method of \texttt{doubly} (Listing \ref{lst:doubly-bug}).
The code uses \texttt{ptr::read} to duplicate ownership of the value from the heap node into \texttt{val}.
It then drops the node by reconstructing a \texttt{Box} from the raw pointer.
However, dropping a \texttt{Box<Node<T>>}(line~8) automatically drops its content.
Since \texttt{ptr::read} duplicates the value without consuming the original,
this drop frees the resource while \texttt{val} still holds a copy, causing a double free.
Our tool detected this by tracking the ownership alias created by \texttt{ptr::read} and flagging the subsequent drop on the aliased path.

\begin{listing}[htbp]
\begin{Verbatim}[commandchars=\\\{\},numbersep=2pt, fontsize=\scriptsize, numbers=left]
\PYG{k}{pub}\PYG{+w}{ }\PYG{k}{fn}\PYG{+w}{ }\PYG{n+nf}{pop\PYGZus{}back}\PYG{p}{(}\PYG{o}{\PYGZam{}}\PYG{k}{mut}\PYG{+w}{ }\PYG{n+nb+bp}{self}\PYG{p}{)}\PYG{+w}{ }\PYG{p}{\PYGZhy{}\PYGZgt{}}\PYG{+w}{ }\PYG{n+nb}{Option}\PYG{o}{\PYGZlt{}}\PYG{n}{T}\PYG{o}{\PYGZgt{}}\PYG{+w}{ }\PYG{p}{\PYGZob{}}
\PYG{+w}{    }\PYG{k}{unsafe}\PYG{+w}{ }\PYG{p}{\PYGZob{}}
\PYG{+w}{        }\PYG{k}{if}\PYG{+w}{ }\PYG{n+nb+bp}{self}\PYG{p}{.}\PYG{n}{length}\PYG{+w}{ }\PYG{o}{=}\PYG{o}{=}\PYG{+w}{ }\PYG{l+m+mi}{0}\PYG{+w}{ }\PYG{p}{\PYGZob{}}
\PYG{+w}{            }\PYG{n+nb}{None}
\PYG{+w}{        }\PYG{p}{\PYGZcb{}}\PYG{+w}{ }\PYG{k}{else}\PYG{+w}{ }\PYG{p}{\PYGZob{}}
\PYG{+w}{            }\PYG{k+kd}{let}\PYG{+w}{ }\PYG{n}{val}\PYG{+w}{ }\PYG{o}{=}\PYG{+w}{ }\PYG{n}{ptr}\PYG{p}{::}\PYG{n}{read}\PYG{p}{(}\PYG{o}{\PYGZam{}}\PYG{p}{(}\PYG{o}{*}\PYG{n+nb+bp}{self}\PYG{p}{.}\PYG{n}{last}\PYG{p}{)}\PYG{p}{.}\PYG{n}{value}\PYG{p}{)}\PYG{p}{;}\PYG{+w}{ }\PYG{c+c1}{// [1] Ownership duplication}
\PYG{+w}{            }\PYG{c+c1}{// ... (pointer updates) ...}
\PYG{+w}{            }\PYG{k+kd}{let}\PYG{+w}{ }\PYG{n}{old\PYGZus{}last}\PYG{+w}{ }\PYG{o}{=}\PYG{+w}{ }\PYG{n+nb+bp}{self}\PYG{p}{.}\PYG{n}{last}\PYG{p}{;}
\PYG{+w}{            }\PYG{c+c1}{// ...}
\PYG{+w}{            }\PYG{n+nb}{drop}\PYG{p}{(}\PYG{n}{mem}\PYG{p}{::}\PYG{n}{transmute}\PYG{p}{:}\PYG{p}{:}\PYG{o}{\PYGZlt{}}\PYG{n}{\PYGZus{}}\PYG{p}{,}\PYG{+w}{ }\PYG{n+nb}{Box}\PYG{o}{\PYGZlt{}}\PYG{n}{Node}\PYG{o}{\PYGZlt{}}\PYG{n}{T}\PYG{o}{\PYGZgt{}}\PYG{o}{\PYGZgt{}}\PYG{o}{\PYGZgt{}}\PYG{p}{(}\PYG{n}{old\PYGZus{}last}\PYG{p}{)}\PYG{p}{)}\PYG{p}{;}\PYG{+w}{ }\PYG{c+c1}{// [2] Drop box implicitly drops value (Free \PYGZsh{}1)}
\PYG{+w}{            }\PYG{n+nb}{Some}\PYG{p}{(}\PYG{n}{val}\PYG{p}{)}\PYG{+w}{ }\PYG{c+c1}{// [3] Caller drop causes Free \PYGZsh{}2}
\PYG{+w}{        }\PYG{p}{\PYGZcb{}}
\PYG{+w}{    }\PYG{p}{\PYGZcb{}}
\PYG{p}{\PYGZcb{}}
\end{Verbatim}

\caption{Ownership Duplication leading to Double Free in \texttt{doubly}.}
\label{lst:doubly-bug}
\end{listing}

\textbf{Case 4: Out-of-Bounds in \texttt{auto\_vec}}.
We identified an Out-of-Bounds (OOB) vulnerability in the iterator implementation of \texttt{auto\_vec}, a vector implementation with automatic child removal.
As shown in Listing \ref{lst:autovec-bug}, the \texttt{iter} method initializes the iterator by calculating a \texttt{current} pointer that points \emph{before} the start of the backing array.
It computes \texttt{(\&children[0]).sub(1)}, intending to increment it back to the start in the first call to \texttt{next()}.
However, in Rust (and LLVM), performing pointer arithmetic that moves a pointer outside the allocated object (even momentarily) is undefined behavior, unless it points one past the end.
Moving \emph{before} the start is always illegal.
Our tool detected this via the \texttt{layout domain}, which tracks the pointer offset relative to the base object.
The operation \texttt{sub(1)} resulted in a negative offset, triggering Rule \textsf{E-OOB/MA}.

\begin{listing}[htbp]
\begin{Verbatim}[commandchars=\\\{\},numbersep=2pt, fontsize=\scriptsize, numbers=left]
\PYG{k}{pub}\PYG{+w}{ }\PYG{k}{fn}\PYG{+w}{ }\PYG{n+nf}{iter}\PYG{p}{(}\PYG{o}{\PYGZam{}}\PYG{n+nb+bp}{self}\PYG{p}{)}\PYG{+w}{ }\PYG{p}{\PYGZhy{}\PYGZgt{}}\PYG{+w}{ }\PYG{n+nc}{Iter}\PYG{o}{\PYGZlt{}}\PYG{n}{T}\PYG{o}{\PYGZgt{}}\PYG{+w}{ }\PYG{p}{\PYGZob{}}
\PYG{+w}{    }\PYG{n}{Iter}\PYG{+w}{ }\PYG{p}{\PYGZob{}}
\PYG{+w}{        }\PYG{n}{last}\PYG{p}{:}\PYG{+w}{ }\PYG{k+kp}{\PYGZam{}}\PYG{n+nc}{self}\PYG{p}{.}\PYG{n}{children}\PYG{p}{[}\PYG{n+nb+bp}{self}\PYG{p}{.}\PYG{n}{len}\PYG{p}{(}\PYG{p}{)}\PYG{+w}{ }\PYG{o}{\PYGZhy{}}\PYG{+w}{ }\PYG{l+m+mi}{1}\PYG{p}{]}\PYG{+w}{ }\PYG{k}{as}\PYG{+w}{ }\PYG{o}{*}\PYG{k}{const}\PYG{+w}{ }\PYG{n}{\PYGZus{}}\PYG{p}{,}
\PYG{+w}{        }\PYG{n}{current}\PYG{p}{:}\PYG{+w}{ }\PYG{n+nc}{unsafe}\PYG{+w}{ }\PYG{p}{\PYGZob{}}
\PYG{+w}{            }\PYG{p}{(}\PYG{o}{\PYGZam{}}\PYG{n+nb+bp}{self}\PYG{p}{.}\PYG{n}{children}\PYG{p}{[}\PYG{l+m+mi}{0}\PYG{p}{]}\PYG{+w}{ }\PYG{k}{as}\PYG{+w}{ }\PYG{o}{*}\PYG{k}{const}\PYG{+w}{ }\PYG{o}{..}\PYG{p}{.}\PYG{p}{)}\PYG{p}{.}\PYG{n}{sub}\PYG{p}{(}\PYG{l+m+mi}{1}\PYG{p}{)}\PYG{+w}{ }\PYG{c+c1}{// [1] UB: moves ptr BEFORE alloc start}
\PYG{+w}{        }\PYG{p}{\PYGZcb{}}\PYG{p}{,}
\PYG{+w}{        }\PYG{n}{lifetime}\PYG{p}{:}\PYG{+w}{ }\PYG{n+nc}{PhantomData}\PYG{p}{,}
\PYG{+w}{    }\PYG{p}{\PYGZcb{}}
\PYG{p}{\PYGZcb{}}
\end{Verbatim}

\caption{Out-of-Bounds access in \texttt{auto\_vec} iterator initialization.}
\label{lst:autovec-bug}
\end{listing}

\section{Discussion}
\label{sec:discussion}

\subsection{Comparison with Baselines}
\label{subsec:comparison-baselines}

%\textbf{Goal and Contract.}
\tool detects potential soundness violations by checking whether unsafe code upholds the safety invariants required by its safe APIs within the modeled scope.
This goal differs from generic bug-finding tools,
which focus on localized execution errors (e.g., panics or crashes)
rather than invalid states escaping through safe boundaries.
Baselines provide strong solutions for specific bug classes.
Rudra is a pattern matcher over unsafe idioms.
SafeDrop focuses on deallocation bugs by recognizing specific lifetime-ending events.
MirChecker applies generic symbolic reasoning over MIR but does not model Rust-specific ownership contracts.
In contrast, \tool restores the semantic context erased by \texttt{unsafe} code.
It maintains a shared state with ownership, object-validity, and layout components.
%\textbf{Cross-Domain Reasoning.}
The value of this design is not that every warning rule uses all components at once,
but that each rule can consume the facts required by the corresponding Rust obligation
within one flow-sensitive MIR state.
Potential soundness violations often span multiple steps and multiple kinds of invariants.
For example, the \texttt{doubly} bug (Listing~\ref{lst:doubly-bug}) starts from duplicating a raw pointer
and later frees the underlying object.
The \texttt{tracing} bug (Listing~\ref{lst:tracing}) performs a move-like action
(\texttt{mem::forget(self)}) and then dereferences pointers derived from \texttt{self}.
These cases require an analysis to track how ownership transfers and drops affect the liveness of aliased pointers.
Pattern-based checks remain valuable for quickly flagging common misuse patterns,
but they are not designed to connect such multi-step interactions.
Similarly, layout-related bugs such as \texttt{wrflib} (Section~\ref{sec:rq1})
require reasoning about derived pointers and bounds across multiple statements.
By tracking these semantics flow-sensitively, \tool can detect complex violations that evade pattern matchers.
Furthermore, to explain potential alarms,
\tool retains alternative object-validity states at control-flow joins instead of collapsing them to a coarse unknown state.
For example, the lifecycle domain represents uncertainty as sets of states (e.g., $\{\texttt{Live}, \texttt{Uninit}\}$),
which helps explain potential alarms but does not eliminate path-insensitivity (Section~\ref{sec:rq2}).

\subsection{Threats to Validity}
\label{subsec:threats}

We identify potential threats to the validity of our experimental results:

\textbf{Internal Validity.}
The accuracy of our evaluation relies on the correctness of our ground truth and verification process.
For Dataset A, we used established CVEs from the RustSec database.
For Dataset B, the lack of ground truth poses a challenge.
To mitigate bias in our manual verification, we use a strict criterion:
we treat a warning as a bug only if we can construct an executable PoC that triggers Undefined Behavior in Miri.
This ensures that our reported True Positives are genuine soundness violations.
Another potential threat is the configuration of baseline tools.
To ensure fairness, we used their latest stable versions with default configurations.
We also minimized libraries to resolve compilation errors, ensuring a fair comparison with baselines.

\textbf{External Validity.}
Our results on Dataset A (46 CVEs) and Dataset B (83 flagged crates) might not generalize to all Rust crates.
Specifically, our strict inclusion criteria for Dataset A resulted in a somewhat imbalanced distribution of bug types,
which may not perfectly reflect the natural distribution of unknown bugs found in the wild (Dataset B).
However, Dataset A still covers the most critical categories of memory safety CVEs (e.g., Uninit, OOB, UAF).
Furthermore, Dataset B is drawn from the entire crates.io registry,
covering diverse domains from low-level utilities to high-level applications.
While we cannot claim effectiveness on non-public codebases,
our evaluation on the public ecosystem provides strong evidence of \tool's real-world applicability.

\subsection{Limitations and Future Work}
\label{subsec:limitations}

\tool has several limitations, though they do not significantly undermine its effectiveness in detecting common vulnerabilities.
\textbf{(1) Scope.}
\tool focuses on sequential memory safety violations and soundness issues.
It does not target application-level logic bugs.
Additionally, it does not model concurrency or synchronization primitives.
As a result, it cannot detect data races or improper use of \texttt{Sync}/\texttt{Send} traits.
However, empirical evidence suggests that sequential memory safety violations remain the primary source of unsoundness in Rust ecosystems.
\textbf{(2) Precision and Over-approximation.}
As a static analyzer, \tool uses conservative warning generation for the modeled cases,
which can introduce false positives.
At the same time, bounded inlining, opaque calls, unmodeled APIs, and lost pointer
identity can still cause false negatives.
Our layout domain uses interval abstraction for pointer arithmetic,
which can be imprecise for complex non-linear computations (e.g., bitwise operations).
Additionally, we handle FFI calls and deep recursion conservatively by over-approximating their effects.
Despite these approximations, our evaluation demonstrates that \tool maintains a practical precision rate.
\textbf{(3) Modeling Scope.}
Our analysis primarily focuses on modeling key APIs that are critical for memory safety semantics,
such as \texttt{Vec}, \texttt{Box}, and raw pointer operations.
We do not fully model the entire Rust standard library, particularly specialized APIs (e.g., custom allocators)
or complex interior mutability patterns (e.g., deep usage of \texttt{RefCell}).
However, our targeted modeling strategy covers the majority of unsafe usage patterns in our datasets,
and unmodeled APIs can be supported with additional engineering effort.

We identify several promising directions for future research.
First, while our framework effectively targets sequential memory safety,
extending it to model concurrency primitives would broaden its applicability to multi-threaded contexts,
such as detecting data races.
Second, integrating more expressive abstract domains could further refine precision
in complex pointer arithmetic involving non-linear constraints.
Finally, as the Rust ecosystem evolves, expanding the semantic modeling of standard library APIs
remains an ongoing engineering effort to support emerging safe abstractions.

\section{Related Work}
\label{sec:related-work}

\smallskip
\noindent
\textbf{Static Analysis for Unsafe Rust.}
Several tools analyze Rust code to detect memory safety risks.
Rupta~\cite{RUPTA2024} performs context-sensitive pointer analysis on Rust MIR,
which provides a foundation for MIR-level reasoning about pointer behavior.
MirChecker~\cite{MIRChecker2021} applies symbolic execution to Rust MIR and targets generic bugs
such as integer overflows and panics.
Rudra~\cite{rudra2021} uses heuristic pattern matching to flag risky unsafe constructs
and reports issues such as incorrect \texttt{Vec::set\_len} usage and variance-related mistakes.
SafeDrop~\cite{SafeDrop2023} focuses on deallocation bugs (e.g., Use-After-Free and Double-Free).
These tools are effective for specific bug classes but do not directly check whether a safe API
upholds Rust's safety invariants.
In contrast, \tool detects potential soundness violations in safe abstractions by modeling
the ownership, object-validity, and layout facts relevant to its warning rules.
Safe4U~\cite{safe4u2025} checks safe encapsulations against contracts mined from API documentation.
We treat it as complementary and exclude it from our baseline comparison.

\smallskip
\noindent
\textbf{Rust Semantics and Verification.}
Stacked Borrows~\cite{StackedBorrows2020} and Tree
Borrows~\cite{TreeBorrows2025} define operational models for Rust aliasing,
borrowing, and pointer provenance.
These semantics are foundational for explaining when reference and pointer
operations are valid.
\tool has a narrower and more implementation-oriented goal: it does not attempt
to implement a complete Rust provenance or aliasing semantics.
Instead, it tracks the ownership, object-validity, and layout facts needed by
its warning rules over MIR, and treats behaviors outside these models, such as
pointer-integer-pointer round trips and unmodeled library semantics, as outside
its current scope.

RustBelt~\cite{RustBelt2018} and Aeneas~\cite{Aeneas2024} provide
proof-oriented foundations and verification workflows for Rust programs.
Such systems target correctness arguments under explicit semantic models and
specifications.
In contrast, \tool is an automatic static bug detector for unsafe
implementations of safe abstractions: it scales to crate-level analysis and
reports potential violations that can be validated dynamically, rather than
proving full Rust semantic correctness.

Place Capability Graphs~\cite{PlaceCapabilityGraphs2025} are closest in spirit
to \tool's ownership reasoning, since they model permissions and capability
changes for Rust places.
The scope differs, however.
PCG provides a general-purpose model of Rust's ownership and borrowing
guarantees, whereas \tool combines ownership facts with object-lifecycle and
layout facts to detect specific classes of memory-safety-relevant violations in
unsafe safe abstractions.
Accordingly, our evaluation measures large-scale bug-finding yield and
validation of reported issues, rather than semantic model coverage or direct
verification strength.

\smallskip
\noindent
\textbf{Dynamic Analysis and Program Verification Tools.}
Miri~\cite{miri} provides a concrete interpreter for detecting Undefined Behavior in Rust
and serves as a ground-truth validator for many analyses, including ours.
Its coverage depends on available test inputs.
Formal tools such as Kani~\cite{trait2024}, Prusti~\cite{Prusti2022}, Creusot~\cite{Creusot2022},
and Verus~\cite{lattuada2023verus} can prove properties against specifications,
but scaling them to large codebases with extensive raw-pointer manipulation remains challenging.
OOM-Guard~\cite{OOM-Guard2023} mitigates out-of-memory risks at runtime and is orthogonal to memory corruption.

\smallskip
\noindent
\textbf{Development Support and Documentation Analysis.}
VRLifeTime~\cite{VRLifeTime2020} and Yuga~\cite{Yuga2023} integrate dataflow analyses into IDEs for lifetime visualization,
primarily targeting safe code and explicit lifetimes.
'R~\cite{R2021} and RustC4~\cite{zhangyichi2024Leveraging} detect inconsistencies between code and documentation,
while rust-code-analysis~\cite{rust-code-analysis2020} provides structural metrics.
Cargo-call-stack~\cite{cargo-call-stack2019} analyzes stack usage for embedded systems.
These tools improve developer workflows but do not directly check soundness invariants inside unsafe implementations.

\section{Conclusion}

In this paper, we presented \tool, a static analysis framework for detecting
potential soundness violations in Rust safe abstractions.
By maintaining ownership, object-validity, and layout facts in a shared MIR-level state,
\tool reports both boundary-level contract violations and instruction-level undefined behavior
within its modeled sequential memory-safety scope.
Our evaluation on a dataset of known vulnerabilities demonstrates that \tool significantly outperforms state-of-the-art tools,
detecting 36 out of 53 ground-truth bugs with 51.6\% alert-level precision.
Additionally, \tool successfully discovered 114 previously unknown bugs in popular crates from crates.io,
with 45 of them confirmed and 27 fixed by developers.

\bibliographystyle{ACM-Reference-Format}
\bibliography{ref.bib}

\appendix

\end{document}